\documentclass[superscriptaddress]{revtex4-1}
\usepackage{amsmath,amsfonts,amssymb,bm}
\newcommand*\Laplace{\mathop{}\!\mathbin\bigtriangleup}
\newcommand*\DAlambert{\mathop{}\!\mathbin\Box}
\numberwithin{equation}{section}
\begin{document}
\title{Path Integrals of the Vector Field\\%
--- {\em Covariance of a Path Integral\/} ---}
\author{Seiji Sakoda}
\email[]{sakoda@nda.ac.jp}
\affiliation{Department of Applied Physics, National Defense Academy,
Hashirimizu\\
Yokosuka city, Kanagawa 239-8686, Japan}
\begin{abstract}
On the basis of the canonical quantization procedure of a system defined on a cubic lattice, we propose a new method, in which resolutions of unity expressed in terms of eigenvectors are naturally provided, to find eigenvectors of field operators. By making use of fundamental ingredients thus obtained, we derive time sliced path integral formulae for a massive vector field accompanied with a scalar field. Due to the indefinite metric of the Hilbert space upon which we define field operators, the action
appears in the path integral looks quite different from the
classical one. Nevertheless we will find that the effective action defined by introducing external sources results in the original action.
By taking the effective action as the base of consideration, we study the proper meaning of the covariance of a path integral.
\end{abstract}
\maketitle

\section{Introduction}
Path integral method\cite{Feynman:48,Feynman:50} is now the most efficient and powerful tool to study the quantum field theory. In particular, since the quantization and the covariant formulation of the non-Abelian gauge theories\cite{Feynman:63,DeWitt:67} was achieved in terms of the path integral by Faddeev and Popov\cite{Faddeev-Popov,'t Hooft:1971NPB33}, the method became widely used as a systematic method for quantizing constrained systems\cite{Faddeev:70,Senjanovic}. In the perturbative treatment of the theory, non-Abelian gauge theory then got great progress by the discovery of the BRS symmetry\cite{BRS} basing on the Faddeev-Popov path integral. Generalized prescription of gauge fixing on the basis of the BRS invariance was then formulated\cite{Kugo-Uehara} after the consistent formulation of the covariant canonical formalism with the use of indefinite metric Hilbert space by Kugo and Ojima\cite{Kugo-Ojima:QB}. On the other hand, in connection with the statistical mechanics\cite{Matsubara,Abrikosov-Grokov-Dzyaloshinski,Poopv:1983,Rivers:1987,Parisi:1988}, the lattice gauge theory\cite{KGWilson:74} opened the way to study non-perturbative aspects of non-Abelian gauge theory in the Euclidean space; lattice formulation of the non-Abelian gauge theory can be viewed as non-perturbative generalization to the continuum formulation of Euclidean path integral\cite{Abers-Lee} of such a system. Even in the continuum representation, the Euclidean technique, such as the instanton calculation\cite{Polyakov:PLB59,Belavin-Polyakov-Schwartz-Tyupkin,'t Hooft:PRL37,Jackiw-Rebbi:PRL37,Callan-Dashen-Gross,Jackiw-Rebbi:PRD14,'t Hooft:PRD14,Shwartz,Jackiw-Rebbi:PLB67,Coleman:85,Atiyah-Hitchin-Singer,Jackiw-Nohl-Rebbi,Atiyah-Hitchin-Drinfeld-Manin}, calculation of anomalies\cite{Fukuda-Miyamoto,Steinberger,Schwinger:PR82,Adler,Bell-Jackiw,Gross-Jackiw,Fujikawa-Anomaly}, etc., often exhibits the usefulness of the path integral method in treating topological objects\cite{Kashiwa-Nima-Sakoda}.
Manifest covariance and being easy to take global aspects of the system into account will be the origin of the powerfulness and the usefulness of the path integral.

On the manifestly covariant path integral of the vector field, on which the Faddeev-Popov procedure\cite{Faddeev-Popov} and the 't Hooft average\cite{'t Hooft:1971NPB33} are formulated, however, there exists a simple question; namely, due to the covariance, quadratic term of the time component in the Lagrangian possesses oposite sign compared to other three spatial components. Nevertheless we always make use of the covariant propagator $\eta_{\mu\nu}/(k^{2}+i\epsilon)$ in perturbative calculations regardless of the vector component ($\eta=\text{diag.}(1,\,-1,\,-1,\,-1)$).
As it is often explained in textbook, $i\epsilon$ term corresponds to the vacuum wave function in the limit of putting the initial and finial times of the Feynman kernel to $\mp\infty$ (see chapter 9 of Ref.~\cite{WeinbergI}, for example). 
The same sign in the denominator of the propagator for all components therefore means that vacuum wave functional of the time component must have the same sign in the quadratic term in the field variable relative to spatial components though from the covariance point of view the time and spatial components are expected to have opposite signs in the quadratic term; the sign change in the time component may therefore indicates the need of indefinite metric representation\cite{Pauli:1943,Henneaux-Teitelboim}.
On this issue, Feynman himself seems to have been aware of already in 1950 according to a comment given inside a footnote in Ref.~\cite{Feynman:50}.
In this regard, Arisue, Fujiwara, Inoue and Ogawa\cite{Arisue:81} formulated, though restricted to the Feynman gauge, a path integral of a massless vector field by making use of the Pauli's generalized Schr\"{o}dinger representation in which the time component of the vector field takes ``{\em imaginary like\/}'' eigenvalue as a consequence of the indefinite metric Hilbert space in covariant canonical formalism. They do not, however, discuss the covariance of their path integral in detail. Soon after Arisue  {\em et al.\/}, Kashiwa\cite{Kashiwa:81} and Kashiwa and Sakamoto\cite{Kashiwa-Sakamoto} presented lattice formulation of the Euclidean path integrals of massive\cite{Kashiwa:81} and massless\cite{Kashiwa-Sakamoto} vector fields by paying attention to the boundary condition of their path integrals to avoid the infrared problem. In their formulation, they seem to make use only of the positive definite Hilbert space to define operators they need to express the Hamiltonian and on this setting they obtain, by introducing dummy variables through Gaussian identities, covariant propagator in Euclidean space. In this connection, it will be interesting to clarify whether there is no need of indefinite metric Hilbert space indeed to get a covariant results by means of the path integral method.
As for the use of indefinite metric representation, the consistency of a path integral for the Lee-Wick model\cite{Lee-Wick:NPB9} was discussed by Boulware and Gross\cite{Boulware-Gross}. On the basis of BRS quantization, Marnelius discussed the reality of the Hamiltonian needed for consistent time evolution when the indefinite metric representation is utilized\cite{Marnelius:PLB318}. In addition to these, from the viewpoint of distributions, Tonder and Dorca argue a wide class of path integrals for phantom and ghost degrees\cite{Tonder-Dorca}. Some of them will be connected to our purpose of this paper but others may beyond our scope.

Another issue we should like to ask is the meaning of the prescription $A_{0}=iA_{4}$\cite{Ramond:1989,Swanson:1992}, i.e. the Wick rotation, like the calculation of anomaly by Fujikawa's method\cite{Fujikawa-Anomaly}, at the same time doing substitution $x^{0}=-ix^{4}$. In Ref.~\cite{Swanson:1992}, in particular, we can find the prescription: $A_{0}\to iA_{0}$, $\pi_{0}\to i\pi_{0}$ and $J_{0}\to iJ_{0}$ at the same time to put $t\to-it$ in order to obtain the Euclidean commutation relations $[A_{\mu},\,\pi_{\nu}]=-i\delta_{\mu,\,\nu}$ and then canonical formalism in Euclidean space is explained. It is difficult to understand the relation of such a theory with the usual covariant canonical formalism in Minkowski space. Usually, naive or pure Euclidean technique will never require such a redefinition of integration variables of the path integral. We shall naively consider the Euclidean path integral in this paper just by introducing the imaginary time and will observe the difference between the result in this situation and the one obtained by the above additional replacement. Since this prescription is now widely accepted, as cited above, including the lattice gauge theory, in formulating the Euclidean path integral of the vector field, the difference, if any, immediately generates a discrepancy between the real gauge theory in the Minkowski space (our Euclidean formulation which transforms to Minkowski result just by putting $x_{4}=ix_{0}$) and the one defined in Euclidean space in terms of the Euclidean vector field introduced by the Wick rotation. The understanding of the relation between these two approach should be therefore of importance from several viewpoints. First of all, some non-perturbative results from the Euclidean formalism such as quark confinement, instanton effects, etc. would depend on the configuration of the Euclidean vector field; we may ask whether we can interpret such phenomena in terms of real vector field defined in Minkowski space? On this point, Arisue {\em et al.\/} state that the meaning of the Wick rotation can be explained by the need of indefinite metric representation in the covariant canonical formalism. If so, we can translate the Euclidean results just by putting $x^{4}=ix^{0}$ without doing Wick rotation of the vector field; it will be then hopeful to understand the results in terms of Minkowski vector field. To do so, we need to generalize the result of Arisue {\em et al.\/} to an arbitrary covariant gauge by removing the restriction to the Feynman gauge. This is our first target of this paper. The second question to be asked will be the connection of the Euclidean covariance in the Euclidean path integral with the one formulated in Minkowski space or, we may rather ask what the covariance of a path integral should be. This is the second aim of this paper.

In this paper we consider quantization of a massive vector field accompanied with Nakanishi-Lautrup $B$-field\cite{Nakanishi:NLB,Lautrup}. We then propose a new systematic method of finding eigenvectors of field operators. Our method involves inside itself a mechanism to yield the resolution or the decomposition of unity, expressed in terms of the eigenvector to be obtained, of the Hilbert space on which the operator under consideration being defined. This is the refinement of Ref.\cite{Aso} by defining field variables on a lattice. We then utilize these resolutions of unity to formulate time sliced Hamiltonian path integrals. Both the Euclidean and the Minkowski path integrals will be considered to investigate the covariance of the path integral. To this aim, we shall calculate the effective action of the system to be the base of consideration on the covariance of the path integral.

The paper is organized as follows: in section 2, we consider the covariant canonical quantization of the system in which we are interested and observe the structure of the Hilbert space on which the system is represented. We then, by taking the metric structure of the Hilbert space into account, construct eigenvectors of field operators. Euclidean path integral will be discussed in section 3. Questions mentioned above will be answered in section 4 by examining the path integral in Minkowski space. The final section is devoted to the conclusion. The appendix will be helpful to understand our idea through a simple quantum mechanical model.

\section{Massive vector field on a lattice}

\subsection{Definition of the system}
We begin with the Lagrangian, proposed by Nakanishi\cite{Nakanishi:72} to view the massless vector field as the massless limit of this model,
\begin{equation}
\label{eq:lag01}
	L=\frac{1}{a}\sum_{\bm{n}=-N}^{N}
	\left\{
	-\frac{1}{4}F^{\mu\nu}{}_{;\,\bm{n}}(x^{0})F_{\mu\nu;\,\bm{n}}(x^{0})+
	\frac{M^{2}}{2}A^{\mu}{}_{;\,\bm{n}}(x^{0})A_{\mu;\,\bm{n}}(x^{0})+
	B_{\bm{n}}(x^{0})\tilde{\nabla}_{\mu}A^{\mu}{}_{;\,\bm{n}}(x^{0})+
	\frac{\alpha}{2}\left(B_{\bm{n}}(x^{0})\right)^{2}\right\}
\end{equation}
in which dimensionless vector filed $A^{\mu}{}_{;\,\bm{n}}(x^{0})$ and $B_{\bm{n}}(x^{0})$ are defined by $A^{\mu}{}_{;\,\bm{n}}\equiv aA^{\mu}(x^{0},\bm{n}a)$ and $B_{\bm{n}}(x^{0})\equiv a^{2}B(x^{0},\bm{n}a)$, respectively, at each lattice point $\bm{n}a$ and $M\equiv ma$ is the dimensionless mass parameter. The lattice spacing $a$ is taken to be $a=L/(2N+1)$ for the cubic lattice with length $L$ and the sum in the above is the abbreviation of sums with respect to $n_{k}=0,\,\pm1,\,\pm2,\,\dots,\,\pm N$ for $k=1,\,2,\,3$. For a field $\phi_{\bm{n}}(x^{0})$ on this lattice, we introduce $\nabla_{\mu}\phi_{\bm{n}}(x^{0})$ and $\tilde{\nabla}_{\mu}\phi_{\bm{n}}(x^{0})$ by
\begin{equation}
\label{eq:nabla01}
	\nabla_{0}\phi_{\bm{n}}(x^{0})=
	\tilde{\nabla}_{0}\phi_{\bm{n}}(x^{0})=
	a\partial_{0}\phi_{\bm{n}}(x^{0}),\quad
	\nabla_{k}\phi_{\bm{n}}(x^{0})=\phi_{\bm{n}}(x^{0})-
	\phi_{\bm{n}-\hat{\bm{k}}}(x^{0}),\
	\tilde{\nabla}_{k}\phi_{\bm{n}}(x^{0})=\phi_{\bm{n+\hat{\bm{k}}}}(x^{0})-
	\phi_{\bm{n}}(x^{0}),
\end{equation}
where $\hat{\bm{k}}$ is a unit vector in the $k$-direction for $k=1,\,2,\,3$. According to this rule, we have defined $F_{\mu\nu;\,\bm{n}}(x^{0})$ above by
\begin{equation}
	F_{\mu\nu;\,\bm{n}}(x^{0})=
	\nabla_{\mu}A_{\nu;\,\bm{n}}(x^{0})-\nabla_{\nu}A_{\mu;\,\bm{n}}(x^{0}).
\end{equation}
We impose the periodic boundary condition (PBC) on this lattice so that the integration by parts, for instance
\begin{equation}
	\sum_{\bm{n}=-N}^{N}\phi^{1}{}_{;\,\bm{n}}(x^{0})
	\nabla_{k}\phi^{2}{}_{;\,\bm{n}}(x^{0})=
	-\sum_{\bm{n}=-N}^{N}\tilde{\nabla}_{k}
	\phi^{1}{}_{;\,\bm{n}}(x^{0})\phi^{2}{}_{;\,\bm{n}}(x^{0}),
\end{equation}
should be free from the surface term.

If we introduce a new vector field $U_{\mu;\,\bm{n}}(x^{0})$ by
\begin{equation}
	U_{\mu;\,\bm{n}}(x^{0})\equiv
	A_{\mu;\,\bm{n}}(x^{0})-\frac{1}{M^{2}}\nabla_{\mu}B_{\bm{n}}(x^{0}),
\end{equation}
the Lagrangian \eqref{eq:lag01} can be rewritten as
\begin{equation}
	\label{eq:lag02}
	L=\frac{1}{a}\sum_{\bm{n}=-N}^{N}
	\left\{
	-\frac{1}{4}G^{\mu\nu}{}_{;\,\bm{n}}(x^{0})G_{\mu\nu;\,\bm{n}}(x^{0})+
	\frac{M^{2}}{2}U^{\mu}{}_{;\,\bm{n}}(x^{0})U_{\mu;\,\bm{n}}(x^{0})-
	\frac{1}{2M^{2}}\nabla_{\mu}B_{\bm{n}}(x^{0})
	\nabla^{\mu}B_{\bm{n}}(x^{0})+
	\frac{\alpha}{2}\left(B_{\bm{n}}(x^{0})\right)^{2}\right\},
\end{equation}
where $G_{\mu\nu;\,\bm{n}}(x^{0})\equiv\nabla_{\mu}U_{\nu;\,\bm{n}}(x^{0})-\nabla_{\nu}U_{\mu;\,\bm{n}}(x^{0})$. We thus observe that our system is equivalent to Proca field $U_{\mu;\,\bm{n}}(x^{0})$ accompanied with a negative normed scalar field $B_{\bm{n}}(x^{0})$ as far as $M^{2}>0$ and $\alpha>0$.

Equations of motion reads
\begin{equation}
\label{eq:eom}
	\left(a^{2}\partial_{0}^{2}-\sum_{k=1}^{3}
	\nabla_{k}\tilde{\nabla}_{k}+M^{2}\right)U_{\mu;\,\bm{n}}(x^{0})=0
\end{equation}
and
\begin{equation}
	\left(a^{2}\partial_{0}^{2}-\sum_{k=1}^{3}
	\nabla_{k}\tilde{\nabla}_{k}+
	\alpha M^{2}\right)B_{\bm{n}}(x^{0})=0.
\end{equation}
If we introduce $F_{n}^{r}$ defined by
\begin{equation}
	F_{n}^{r}=\frac{1}{\sqrt{2N+1}}
	\exp\left(\frac{2\pi irn}{2N+1}\right),
\end{equation}
we can expand $U_{\mu;\,\bm{n}}(x^{0})$ into a series as
\begin{equation}
	U_{\mu;\,\bm{n}}(x^{0})=\sum_{\bm{r}=-N}^{N}F_{\bm{n}}^{\bm{r}}
	\tilde{U}_{\mu;\,\bm{r}}(x^{0}),
\end{equation}
where $F_{\bm{n}}^{\bm{r}}\equiv F_{n_{1}}^{r_{1}}F_{n_{2}}^{r_{2}}F_{n_{3}}^{r_{3}}$.
Inversion of the above expansion is immediate by making use of identities
\begin{equation}
	\sum_{\bm{n}=-N}^{N}F_{\bm{n}}^{\bm{r}}F_{\bm{n}}^{\bm{r}{'}*}=
	\delta_{\bm{r},\,\bm{r}{'}},\quad
	\sum_{\bm{r}=-N}^{N}F_{\bm{n}}^{\bm{r}*}F_{\bm{n}{'}}^{\bm{r}}=
	\delta_{\bm{n},\,\bm{n}{'}},
\end{equation}
where $\delta_{\bm{n},\,\bm{n}{'}}\equiv\delta_{n_{1},\,n_{1}{'}}\delta_{n_{2},\,n_{2}{'}}\delta_{n_{3},\,n_{3}{'}}$.
The action of the Laplacian $\Laplace\equiv\sum_{k=1}^{3}\nabla_{k}\tilde{\nabla}_{k}$ on $F_{\bm{n}}^{\bm{r}}$ yields
\begin{equation}
	\Laplace F_{\bm{n}}^{\bm{r}}=
	-2\sum_{k=1}^{3}\left\{1-\cos\left(\frac{2\pi r_{k}}{2N+1}\right)\right\}
	F_{\bm{n}}^{\bm{r}}
\end{equation}
so that we define
\begin{equation}
	\omega_{\bm{r}}\equiv\sqrt{m^{2}+p^{2}_{\bm{r}}},\quad
	p^{2}_{\bm{r}}\equiv\frac{2}{a^{2}}
	\sum_{k=1}^{3}\left\{1-\cos\left(\frac{2\pi r_{k}}{2N+1}\right)\right\},
\end{equation}
to be able to solve \eqref{eq:eom} as
\begin{equation}
\label{eq:cl_sol}
	U_{\mu;\,\bm{n}}(x^{0})=\sum_{\bm{r}=-N}^{N}
	\frac{1}{\sqrt{2a\omega_{\bm{r}}}}\left(
	a_{\mu;\,\bm{r}}e^{-i\omega_{\bm{r}}x^{0}}F_{\bm{n}}^{\bm{r}}+
	a^{*}_{\mu;\,\bm{r}}e^{i\omega_{\bm{r}}x^{0}}F_{\bm{n}}^{\bm{r}*}\right).
\end{equation}
In the same way $B_{\bm{n}}(x^{0})$ can be expressed as
\begin{equation}
	B_{\bm{n}}(x^{0})=\sum_{\bm{r}=-N}^{N}
	\frac{M}{\sqrt{2a\omega{'}_{\bm{r}}}}\left(
	b_{\bm{r}}e^{-i\omega{'}_{\bm{r}}x^{0}}F_{\bm{n}}^{\bm{r}}+
	b^{*}_{\bm{r}}e^{i\omega{'}_{\bm{r}}x^{0}}F_{\bm{n}}^{\bm{r}*}\right),
\end{equation}
where we have defined $\omega{'}_{\bm{r}}$ by replacing $m^{2}$ with $\alpha m^{2}$ in $\omega_{\bm{r}}$.

To achieve the canonical quantization, we choose $U^{\mu}{}_{;\,\bm{n}}$ and $B_{\bm{n}}(x^{0})$ as independent variables and introduce their canonical conjugates by
\begin{equation}
	\varPi_{0;\,\bm{n}}(x^{0})=0,\quad
	\varPi_{k;\,\bm{n}}(x^{0})=\nabla_{0}U_{k;\,\bm{n}}(x^{0})-
	\nabla_{k}U_{0;\,\bm{n}}(x^{0}),\quad
	\varPi_{B;\,\bm{n}}(x^{0})=-\frac{1}{M^{2}}\nabla_{0}B_{\bm{n}}(x^{0}),
\end{equation}
where the primary constraint $\phi^{1}{}_{;\,\bm{n}}(x^{0})\equiv\varPi_{0}{}_{;\,\bm{n}}(x^{0})=0$ leads to the secondary constraint $\phi^{2}{}_{;\,\bm{n}}(x^{0})\equiv\sum_{k=1}^{3}\tilde{\nabla}_{k}\varPi_{k;\,\bm{n}}(x^{0})+M^{2}U_{0;\,\bm{n}}(x^{0})=0$ and they form a pair of second class constraints.
We may therefore eliminate $U_{0;\,\bm{n}}(x^{0})$ and $\varPi_{0;\,\bm{n}}(x^{0})$ by setting $U_{0;\,\bm{n}}(x^{0})=-\sum_{k=1}^{3}\tilde{\nabla}_{k}\varPi_{k;\,\bm{n}}/M^{2}$ and $\varPi_{0;\,\bm{n}}(x^{0})=0$, respectively.
Then the quantum Hamiltonian reads
\begin{multline}
\label{eq:ham01}
	\hat{H}=\frac{1}{a}\sum_{\bm{n}=-N}^{N}\left[
	\frac{1}{2}\sum_{k=1}^{3}\left\{
	\left(\hat{\varPi}_{k}{}_{;\,\bm{n}}(x^{0})\right)^{2}
	+M^{2}\left(\hat{U}_{k;\,\bm{n}}(x^{0})\right)^{2}\right\}+
	\frac{1}{2M^{2}}\left(
	\sum_{k=1}^{3}\tilde{\nabla}_{k}\hat{\varPi}_{k;\,\bm{n}}\right)^{2}
	+\frac{1}{4}\sum_{k,l=1}^{3}
	\hat{G}^{k\,l}{}_{;\,\bm{n}}\hat{G}_{k\,l;\,\bm{n}}\right.\\-\left.
	\frac{1}{2}\left\{M^{2}\left(\hat{\varPi}_{B;\,\bm{n}}(x^{0})\right)^{2}-
	\frac{1}{M^{2}}\hat{B}_{\bm{n}}(x^{0})\Laplace
	\hat{B}_{\bm{n}}(x^{0})+
	\alpha\left(\hat{B}_{\bm{n}}(x^{0})\right)^{2}\right\}\right].
\end{multline}
Canonical commutation relations are given by
\begin{equation}
	[\hat{U}_{k;\,\bm{n}}(x^{0}),\,
	\hat{\varPi}_{l;\,\bm{n}{'}}(x^{0})]
	=i\delta_{k,\,l}
	\delta_{\bm{n},\,\bm{n}{'}},\quad
	[\hat{B}_{\bm{n}}(x_{0}),\,\hat{\varPi}_{B;\,\bm{n}{'}}(x^{0})]=
	i\delta_{\bm{n},\,\bm{n}{'}}
\end{equation}
in addition to other trivial ones. In terms of $\hat{P}_{k;\,\bm{n}}(x^{0})\equiv\nabla_{0}\hat{U}_{k;\,\bm{n}}(x^{0})$ and $P_{B;\,\bm{n}}(x^{0})\equiv\nabla_{0}B_{\bm{n}}(x^{0})$ instead of $\hat{\varPi}_{k;\,\bm{n}}(x^{0})$ and $\hat{\varPi}_{B;\,\bm{n}}(x^{0})$ above, we can express the commutation relations as
\begin{equation}
	[\hat{U}_{k;\,\bm{n}}(x^{0}),\,\hat{P}_{l;\,\bm{n}{'}}(x^{0})]=
	i\left(\delta_{k,\,l}-\frac{\tilde{\nabla}_{k}\nabla_{l}}{M^{2}}\right)
	\delta_{\bm{n},\,\bm{n}{'}},\quad
	[\hat{B}_{\bm{n}}(x_{0}),\,\hat{P}_{B;\,\bm{n}{'}}(x^{0})]=
	-iM^{2}\delta_{\bm{n},\,\bm{n}{'}}.
\end{equation}
The classical solution in \eqref{eq:cl_sol} is now converted into the corresponding Heisenberg operator as
\begin{equation}
\label{eq:qm_sol}
	\hat{U}_{k;\,\bm{n}}(x^{0})=\sum_{\bm{r}=-N}^{N}
	\frac{1}{\sqrt{2a\omega_{\bm{r}}}}\left(
	\hat{a}_{k;\,\bm{r}}
	e^{-i\omega_{\bm{r}}x^{0}}F_{\bm{n}}^{\bm{r}}+
	\hat{a}^{\dagger}_{k;\,\bm{r}}
	e^{i\omega_{\bm{r}}x^{0}}F_{\bm{n}}^{\bm{r}*}\right).
\end{equation}
In the same way, we may put
\begin{equation}
	\hat{B}_{\bm{n}}(x^{0})=\sum_{\bm{r}=-N}^{N}
	\frac{M}{\sqrt{2a\omega{'}_{\bm{r}}}}\left(
	\hat{b}_{\bm{r}}e^{-i\omega{'}_{\bm{r}}x^{0}}F_{\bm{n}}^{\bm{r}}+
	\hat{b}^{\dagger}_{\bm{r}}e^{i\omega{'}_{\bm{r}}x^{0}}F_{\bm{n}}^{\bm{r}*}
	\right).
\end{equation}
We solve the expansions above to find
\begin{equation}
	\hat{a}_{k}{}_{;\,\bm{r}}e^{-i\omega_{\bm{r}}x^{0}}
	=\sum_{\bm{n}=-N}^{N}
	\sqrt{\frac{a\omega_{\bm{r}}}{2}}F_{\bm{n}}^{\bm{r}*}\left(
	\hat{U}_{k;\,\bm{n}}(x^{0})+\frac{i}{a\omega_{\bm{r}}}
	\hat{P}_{k;\,\bm{n}}(x^{0})\right)
\end{equation}
as well as
\begin{equation}
	\hat{b}_{\bm{r}}e^{-i\omega{'}_{\bm{r}}x^{0}}
	=\sum_{\bm{n}=-N}^{N}
	\sqrt{\frac{a\omega{'}_{\bm{r}}}{2M^{2}}}F_{\bm{n}}^{\bm{r}*}\left(
	\hat{B}_{\bm{n}}(x^{0})+\frac{i}{a\omega{'}_{\bm{r}}}
	\hat{P}_{B;\,\bm{n}}(x^{0})\right).
\end{equation}

The Schr\"{o}dinger operators are obtained by setting $x^{0}=0$ above and there holds
\begin{equation}
	[\hat{a}_{k;\,\bm{r}},\,\hat{a}^{\dagger}_{l;\,\bm{r}{'}}]=
	\delta^{\bm{r},\,\bm{r}{'}}\left(
	\delta_{k,\,l}+
	\frac{P_{k;\,\bm{r}}P^{*}_{l;\,\bm{r}}}{M^{2}}\right),\quad
	[\hat{b}_{\bm{r}},\,\hat{b}^{\dagger}_{\bm{r}{'}}]=
	-\delta_{\bm{r},\,\bm{r}{'}},
\end{equation}
where we have written
\begin{equation}
	\nabla_{k}F_{\bm{n}}^{\bm{r}}=iP_{k;\,\bm{r}}F_{\bm{n}}^{\bm{r}},\quad
	\tilde{\nabla}_{k}F_{\bm{n}}^{\bm{r}}=
	iP^{*}_{k;\,\bm{r}}F_{\bm{n}}^{\bm{r}},\quad
	P_{k;\,\bm{r}}\equiv2e^{-i\pi r_{k}/(2N+1)}
	\sin\left(\frac{\pi r_{k}}{2N+1}\right).
\end{equation}
Note that there holds
\begin{equation}
	\sum_{k=1}^{3}P^{*}_{k;\,\bm{r}}P_{k;\,\bm{r}}=
	P^{2}_{\bm{r}},
\end{equation}
where $P^{2}_{\bm{r}}\equiv a^{2}p^{2}_{\bm{r}}$ and we can diagonalize the matrix $\delta_{k,\,l}+P_{k;\,\bm{r}}P^{*}_{l;\,\bm{r}}/M^{2}$ to be $\text{diag}(1,\,1,\,(P^{2}_{\bm{r}}+M^{2})/M^{2})$.

The vacuum is defined to be destroyed by all $\hat{a}_{k;\,\bm{r}}$ and $\hat{b}_{\bm{r}}$ such that $\hat{a}_{k;\,\bm{r}}\vert0\rangle=\hat{b}_{\bm{r}}\vert0\rangle=0$ in addition to the assumption $\langle0\vert0\rangle=1$. State vectors in the Fock representation are obtained by multiplying $\hat{a}^{\dagger}_{k;\,\bm{r}}$ and $\hat{b}^{\dagger}$ to the vacuum. To see the structure of the Fock space, we may for a while redefine $\hat{a}_{k;\,\bm{r}}$ by unitary transformation and scaling so that they fulfill $[\hat{a}_{k;\,\bm{r}},\hat{a}^{\dagger}_{l;\,\bm{r}{'}}]=\delta_{k,\,l}\delta_{\bm{r},\,\bm{r}{'}}$. If we write, in this new definition,
\begin{equation}
	\vert\{m\}\rangle\equiv
	\prod_{\bm{r}=-N}^{N}
	\frac{1}{\sqrt{(m_{B;\,\bm{r}})!\prod_{k=1}^{3}(m_{k;\,\bm{r}})!}}
	(\hat{b}^{\dagger}_{\bm{r}})^{m_{B;\,\bm{r}}}
	\prod_{k=1}^{3}
	(\hat{a}^{\dagger}_{k;\,\bm{r}})^{m_{k;\,\bm{r}}}\vert0\rangle,
\end{equation}
we obtain
\begin{equation}
	\langle\{m\}\vert\{m{'}\}\rangle=\prod_{\bm{r}=-N}^{N}
	(-1)^{m_{B;\,\bm{r}}}
	\delta_{m_{B;\,\bm{r}},\,m{'}_{B;\,\bm{r}}}
	\prod_{k=1}^{3}\delta_{m_{k;\,\bm{r}},\,m{'}_{k;\,\bm{r}}},
\end{equation}
where $\prod_{\bm{r}=-N}^{N}$ abbreviates $\prod_{r_{1}=-N}^{N}\prod_{r_{2}=-N}^{N}\prod_{r_{3}=-N}^{N}$.
We may therefore define
\begin{equation}
	\langle\underline{\{m\}}\vert=\langle0\vert
	\prod_{\bm{r}=-N}^{N}
	\frac{(-1)^{m_{B;\,\bm{r}}}(\hat{b}_{\bm{r}})^{m_{B;\,\bm{r}}}}
	{\sqrt{(m_{B;\,\bm{r}})!\prod_{k=1}^{3}(m_{k;\,\bm{r}})!}}
	\prod_{k=1}^{3}
	(\hat{a}_{k;\,\bm{r}})^{m_{k;\,\bm{r}}}
\end{equation}
as the conjugate for $\vert\{m\}\rangle$ so that we can express the resolution of unity as
\begin{equation}
	\sum_{\{m\}=0}^{\infty}\vert\{m\}\rangle\langle\underline{\{m\}}\vert
	=\bm{1},
\end{equation}
where the sum should be taken for all $m_{k;\,\bm{r}}$s in addition to $m_{B;\,\bm{r}}$s. Since we have confirmed the metric structure of the Fock space, we now return to the original definition of $\hat{a}_{k;\,\bm{r}}$ and proceed to consider the eigenvectors of the field operators.

\subsection{Eigenvectors of field operators}
To formulate path integrals for a system, the fundamental ingredients we need is the resolution of unity. When we utilize the holomorphic representation, the resolution of unity will be expressed in terms of the coherent state. We here try to construst eigenvectors of field operators and shall find suitable expressions for the resolution of unity on the Hilbert space equipped with the indefinite metric as we have senn in the previous subsection.
 
To obtain the field diagonal representation, let us introduce for spatial components of the vector field
\begin{equation}
	\hat{U}^{(+)}_{k;\,\bm{n}}\equiv
	\sum_{\bm{r}=-N}^{N}
	\frac{1}{\sqrt{2a\omega_{\bm{r}}}}
	\hat{a}_{k;\,\bm{r}}
	F_{\bm{n}}^{\bm{r}},\quad
	\hat{U}^{(-)}_{k;\,\bm{n}}\equiv
	\sum_{\bm{r}=-N}^{N}
	\frac{1}{\sqrt{2a\omega_{\bm{r}}}}
	\hat{a}^{\dagger}_{k;\,\bm{r}}
	F_{\bm{n}}^{\bm{r}*}.
\end{equation}
Their commutation relation reads
\begin{equation}
	[\hat{U}^{(+)}_{k;\,\bm{n}},\,\hat{U}^{(-)}_{l;\,\bm{n}{'}}]=
	\frac{1}{2}\left(K^{-1}\right)^{k,\,l}_{\bm{n},\,\bm{n}{'}},\quad
	\left(K^{-1}\right)^{k,\,l}_{\bm{n},\,\bm{n}{'}}\equiv
	\sum_{\bm{r}=-N}^{N}\frac{1}{a\omega_{\bm{r}}}
	\left(\delta_{k,\,l}+
	\frac{P_{k;\,\bm{r}}P^{*}_{l;\,\bm{r}}}{M^{2}}\right)
	F_{\bm{n}}^{\bm{r}}F_{\bm{n}{'}}^{\bm{r}*}.
\end{equation}
By regarding these operators as creation and annihilation operators, we make use of the technique shown in the appendix to find eigenvectors of $\hat{U}_{k;\,\bm{n}}$ in the following.

We define the coherent state $\vert\{U^{(+)}_{k}\}\rangle$ given by
\begin{equation}
	\vert\{U^{(+)}_{k}\}\rangle=\exp\left\{
	2\sum_{\bm{n},\,\bm{n}{'}=-N}^{N}\sum_{k,\,l=1}^{3}
	\hat{U}^{(-)}_{k;\,\bm{n}}K^{k,\,l}_{\bm{n},\,\bm{n}{'}}
	U^{(+)}_{l;\,\bm{n}{'}}\right\}
	\vert0\rangle
\end{equation}
to be an eigenvector of $\hat{U}^{(+)}_{k;\,\bm{n}}$;
\begin{equation}
	\hat{U}^{(+)}_{k;\,\bm{n}}\vert\{U^{(+)}_{k}\}\rangle=
	\vert\{U^{(+)}_{k}\}\rangle U^{(+)}_{k;\,\bm{n}}.
\end{equation}
If we write $(\vert\{U^{(+)}_{k}\}\rangle)^{\dagger}$ as $\langle\{U^{(-)}_{k}\}\vert$, there holds
\begin{equation}
	\langle\{U^{(-)}_{k}\}\vert\hat{U}^{(-)}_{k;\,\bm{n}}=
	U^{(-)}_{k;\,\bm{n}}\langle\{U^{(-)}_{k}\}\vert,
\end{equation}
where $U^{(-)}_{k;\,\bm{n}}=\left(U^{(+)}_{k;\,\bm{n}}\right)^{*}$.
The inner product of these coherent state is given by
\begin{equation}
	\langle\{U^{(-)}_{k}\}\vert\{U{'}^{(+)}_{k}\}\rangle=
	\exp\left\{2\sum_{\bm{n}=-N}^{N}\sum_{k,\,l=1}^{3}
	U^{k(-)}_{k;\,\bm{n}{'}}K^{k,\,l}_{\bm{n},\,\bm{n}{'}}
	U{'}^{(+)}_{l;\,\bm{n}{'}}\right\}.
\end{equation}
We can check that the coherent state forms complete set in the Fock space defined above for spatial components of the vector field. Furthermore, since operators $\hat{U}_{k;\,\bm{n}}$ and $\hat{\varPi}_{l;\,\bm{n}}$ can be expressed as linear combinations of $\hat{U}^{(\pm)}_{k;\,\bm{n}}$ above, we can formulate a time sliced Hamiltonian path integral by making use of the resolution of unity in terms of the coherent state. Our aim of this paper is, however, to examine properties of path integrals defined by means of eigenvectors of these field operators. So we here omit the detail. (See the end of the appendix for an example of the use of coherent states in formulating path integrals.)

We consider the operator   
\begin{equation}
	\exp\left(i\sum_{\bm{n}=-N}^{N}\sum_{k=1}^{3}
	\lambda_{k;\,\bm{n}}\hat{U}_{k;\,\bm{n}}\right)
\end{equation}
and decompose it into the product
\begin{equation}
	\exp\left(i\sum_{\bm{n}=-N}^{N}\sum_{k=1}^{3}
	\lambda_{k;\,\bm{n}}\hat{U}^{(-)}_{k;\,\bm{n}}\right)
	\exp\left(i\sum_{\bm{n}=-N}^{N}\sum_{k=1}^{3}
	\lambda_{k;\,\bm{n}}\hat{U}^{(+)}_{k;\,\bm{n}}\right)
	\exp\left(-\frac{1}{4}\sum_{\bm{n}=-N}^{N}\sum_{k,\,l=1}^{3}
	\lambda_{k;\,\bm{n}}
	\left(K^{-1}\right)^{k,\,l}_{\bm{n},\,\bm{n}{'}}
	\lambda_{l;\,\bm{n}{'}}\right)
\end{equation}
to find
\begin{equation}
\begin{aligned}
	&\int_{-\infty}^{\infty}\!\!\prod_{\bm{n}=-N}^{N}\prod_{k=1}^{3}
	\frac{d\lambda_{k;\,\bm{n}}}{2\pi}\langle\{U^{(-)}_{k}\}\vert
	\exp\left\{i\sum_{\bm{n}=-N}^{N}\sum_{k=1}^{3}
	\lambda_{k;\,\bm{n}}\left(\hat{U}_{k;\,\bm{n}}-
	U_{k;\,\bm{n}}\right)\right\}\vert\{U{'}^{(+)_{k}}\}\rangle\\
	=&
	\int_{-\infty}^{\infty}\!\!\prod_{\bm{n}=-N}^{N}\prod_{k=1}^{3}
	\frac{d\lambda_{k;\,\bm{n}}}{2\pi}
	\exp\left\{-\frac{1}{4}\sum_{\bm{n},\,\bm{n}{'}=-N}^{N}\sum_{k,\,l=1}^{3}
	\lambda_{k;\,\bm{n}}
	\left(K^{-1}\right)^{k,\,l}_{\bm{n},\,\bm{n}{'}}
	\lambda_{l;\,\bm{n}{'}}-
	i\sum_{\bm{n}=-N}^{N}\sum_{k=1}^{3}
	\lambda_{k;\,\bm{n}}\left(
	U_{k;\,\bm{n}}-
	U^{(-)}_{k;\,\bm{n}}-U{'}^{(+)}_{k;\,\bm{n}}\right)
	\right\}\\
	&\times
	\langle\{U^{(-)}_{k}\}\vert\{U{'}^{(+)}_{k}\}\rangle,
\end{aligned}
\end{equation}
where $U_{k;\,\bm{n}}$s are real valued $c$-number functions.

By integrating $\lambda_{k;\,\bm{n}}$s, we obtain
\begin{equation}
\label{eq:projector01}
	\begin{aligned}
	&\int_{-\infty}^{\infty}\!\!\prod_{\bm{n}=-N}^{N}\prod_{k=1}^{3}
	\frac{d\lambda_{k;\,\bm{n}}}{2\pi}\langle\{U^{(-)}_{k}\}\vert
	\exp\left\{i\sum_{\bm{n}=-N}^{N}\sum_{k=1}^{3}
	\lambda_{k;\,\bm{n}}\left(\hat{U}_{k;\,\bm{n}}-
	U_{k;\,\bm{n}}\right)\right\}\vert\{U{'}^{(+)}_{k}\}\rangle\\
	=&
	\frac{1}{\sqrt{\det\left(\pi K^{-1}\right)}}
	\exp\left\{-\sum_{\bm{n},\,\bm{n}{'}}\sum_{k,\,l=1}^{3}
	\left(U_{k;\,\bm{n}}-
	U^{(-)}_{k;\,\bm{n}}-U{'}^{(+)}_{k;\,\bm{n}}\right)
	K^{k,\,l}_{\bm{n},\,\bm{n}{'}}\left(
	U_{l;\,\bm{n}}-
	U^{(-)}_{l;\,\bm{n}}-U{'}^{(+)}_{l;\,\bm{n}}\right)\right\}
	\langle\{U^{(-)}_{k}\}\vert\{U{'}^{(+)}_{k}\}\rangle,
\end{aligned}
\end{equation}
where $\det(\pi K^{-1})$ being given by
\begin{equation}
	\det(\pi K^{-1})=\prod_{\bm{r}=-N}^{N}\left\{\left(
	\frac{\pi}{a\omega_{\bm{r}}}\right)^{3}
	\frac{P^{2}_{\bm{r}}+M^{2}}{M^{2}}\right\}.
\end{equation}

If we notice that $U{'}^{(+)}_{k;\,\bm{n}}$ and $U^{(-)}_{k;\,\bm{n}}$ are nothing but eigenvalues of corresponding operators, we can further rewrite it as
\begin{equation}
	\langle\{U^{(-)}_{k}\}\vert\{U_{k}\}\rangle
	\langle\{U_{k}\}\vert\{U{'}^{(+)}_{k}\}\rangle,
\end{equation}
where
\begin{equation}
	\vert\{U_{k}\}\rangle\equiv
	\frac{1}{\det\left(\pi K^{-1}\right)^{1/4}}
	\exp\left\{-\sum_{\bm{n},\,\bm{n}{'}}\sum_{k,\,l=1}^{3}\left(
	\frac{1}{2}U_{k;\,\bm{n}}K^{k,\,l}_{\bm{n},\,\bm{n}{'}}
	U_{l;\,\bm{n}{'}}-
	2U_{k;\,\bm{n}}K^{k\,l}_{\bm{n},\,\bm{n}{'}}
	\hat{U}^{(-)}_{k;\,\bm{n}{'}}+
	\hat{U}^{(-)}_{k;\,\bm{n}}K^{k,\,l}_{\bm{n},\,\bm{n}{'}}
	\hat{U}^{(-)}_{l;\,\bm{n}{'}}\right)\right\}
	\vert0\rangle
\end{equation}
and
\begin{equation}
	\langle\{U_{k}\}\vert=
	\frac{1}{\det\left(\pi K^{-1}\right)^{1/4}}
	\langle0\vert
	\exp\left\{-\sum_{\bm{n},\,\bm{n}{'}}\sum_{k,\,l=1}^{3}\left(
	\frac{1}{2}U_{k;\,\bm{n}}K^{k,\,l}_{\bm{n},\,\bm{n}{'}}
	U_{l;\,\bm{n}{'}}-
	2U_{k;\,\bm{n}}K^{k,\,l}_{\bm{n},\,\bm{n}{'}}
	\hat{U}^{(+)}_{k;\,\bm{n}{'}}+
	\hat{U}^{(+)}_{k;\,\bm{n}}K^{k,\,l}_{\bm{n},\,\bm{n}{'}}
	\hat{U}^{(+)}_{l;\,\bm{n}{'}}\right)\right\}.
\end{equation}

If we carry out the Gaussian integrals with respect to $U_{k;\,\bm{n}}$s in the right hand side of \eqref{eq:projector01}, we immediately recognize that
\begin{equation}
\label{eq:resU}
	\int_{-\infty}^{\infty}\!\!\prod_{\bm{n}=-N}^{N}\prod_{k=1}^{3}
	dU_{k;\,\bm{n}}
	\vert\{U_{k}\}\rangle\langle\{U_{k}\}\vert=\bm{1}
\end{equation}
holds as the resolution of unity on the Fock space of spatial components.
We can also check that $\vert\{U_{k}\}\rangle$ and $\langle\{U_{k}\}\vert$ are right and left eigenvectors of $\hat{U}_{k;\,\bm{n}}$;
\begin{equation}
	\hat{U}_{k;\,\bm{n}}\vert\{U_{k}\}\rangle=
	\vert\{U_{k}\}\rangle U_{k;\,\bm{n}},\quad
	\langle\{U_{k}\}\vert\hat{U}_{k;\,\bm{n}}=
	U_{k;\,\bm{n}}\langle\{U_{k}\}\vert.
\end{equation}
The inner product of these eigenvectors is given by
\begin{equation}
	\langle\{U_{k}\}\vert\{U{'}_{k}\}\rangle=
	\prod_{\bm{n}=-N}^{N}\prod_{k=1}^{3}\delta(U_{k;\,\bm{n}}-
	U{'}_{k;\,\bm{n}}).
\end{equation}

Decomposition of $\hat{P}_{k;\,\bm{n}}$ into positive and negative frequency parts
\begin{equation}
	\hat{P}^{(+)}_{k;\,\bm{n}}\equiv-i
	\sum_{\bm{r}=-N}^{N}
	\sqrt{\frac{a\omega_{\bm{r}}}{2}}
	\hat{a}_{k;\,\bm{r}}
	F_{\bm{n}}^{\bm{r}},\quad
	\hat{P}^{(-)}_{k;\,\bm{n}}\equiv i
	\sum_{\bm{r}=-N}^{N}
	\sqrt{\frac{a\omega_{\bm{r}}}{2}}
	\hat{a}^{\dagger}_{k;\,\bm{r}}
	F_{\bm{n}}^{\bm{r}*}
\end{equation}
yields the commutation relations
\begin{equation}
	[\hat{P}^{(+)}_{k;\,\bm{n}},\,\hat{P}^{(-)}{}_{l;\,\bm{n}{'}}]=
	\frac{1}{2}\tilde{K}^{k,\,l}_{\bm{n},\,\bm{n}{'}},\quad
	\tilde{K}^{k,\,l}_{\bm{n},\,\bm{n}{'}}\equiv
	\sum_{\bm{r}=-N}^{N}a\omega_{\bm{r}}\left(
	\delta_{k,\,l}+\frac{P_{k;\,\bm{r}}P^{*}_{l;\,\bm{r}}}{M^{2}}
	\right)F_{\bm{n}}^{\bm{r}}F_{\bm{n}{'}}^{\bm{r}*}.
\end{equation}
We then define the coherent state
\begin{equation}
	\vert\{P^{(+)}_{k}\}\rangle=\exp\left\{
	2\sum_{\bm{n},\,\bm{n}{'}=-N}^{N}\sum_{k,\,l=1}^{3}
	\hat{P}^{(-)}_{k;\,\bm{n}}
	\left(\tilde{K}^{-1}\right)^{k,\,l}_{\bm{n},\,\bm{n}{'}}
	P^{(+)}_{l;\,\bm{n}{'}}\right\}
	\vert0\rangle
\end{equation}
to be an eigenvector of $\hat{P}^{(+)}_{k;\,\bm{n}}$;
\begin{equation}
	\hat{P}^{(+)}_{k;\,\bm{n}}\vert\{P^{(+)}_{k}\}\rangle=
	\vert\{P^{(+)}_{k}\}\rangle P^{(+)}_{k;\,\bm{n}}.
\end{equation}
By writing $(\vert\{P^{(+)}_{k}\}\rangle)^{\dagger}$ as $\langle\{P^{k(-)}\}\vert$, we find
\begin{equation}
	\langle\{P^{(-)}_{k}\}\vert\hat{P}^{(-)}_{k;\,\bm{n}}=
	P^{(-)}_{k;\,\bm{n}}\langle\{P^{(-)}_{k}\}\vert,
\end{equation}
where $P^{(-)}_{k;\,\bm{n}}=\left(P^{(+)}_{k;\,\bm{n}}\right)^{*}$.
The inner product of these coherent state reads
\begin{equation}
	\langle\{P^{(-)}_{k}\}\vert\{P{'}^{(+)}_{k}\}\rangle=
	\exp\left\{2\sum_{\bm{n},\,\bm{n}{'}=-N}^{N}\sum_{k=1}^{3}
	P^{(-)}_{k;\,\bm{n}{'}}
	\left(\tilde{K}^{-1}\right)^{k,\,l}_{\bm{n},\,\bm{n}{'}}
	P{'}^{(+)}_{l;\,\bm{n}{'}}\right\}.
\end{equation}

Eigenvectors of $\hat{P}^{(-)}_{k;\,\bm{n}}$ can be obtained through the definition of the projection operator
\begin{equation}
	\vert\{P_{k}\}\rangle\langle\{P_{k}\}\vert\equiv
	\int\!\!\prod_{\bm{n}=-N}^{N}\prod_{k=1}^{3}
	\frac{d\lambda_{k;\,\bm{n}}}{2\pi}
	\exp\left\{i\sum_{\bm{n}=-N}^{N}\sum_{k=1}^{3}
	\lambda_{k;\,\bm{n}}\left(
	\hat{P}_{k;\,\bm{n}}-P_{k;\,\bm{n}}\right)\right\},
\end{equation}
where $P_{k;\,\bm{n}}$s are real valued $c$-number functions.
Repeating the similar procedure for the case of $\vert\{U_{k}\}\rangle\langle\{U_{k}\}\vert$ above, we obtain
\begin{multline}
	\vert\{P_{k}\}\rangle=
	\frac{1}{\det\left(\pi\tilde{K}\right)^{1/4}}\\
	\times
	\exp\left\{-\sum_{\bm{n},\,\bm{n}{'}}\sum_{k,\,l=1}^{3}
	\left(
	\frac{1}{2}P_{k;\,\bm{n}}
	\left(\tilde{K}^{-1}\right)^{k,\,l}_{\bm{n},\,\bm{n}{'}}
	P_{l;\,\bm{n}{'}}-
	2P_{k;\,\bm{n}}
	\left(\tilde{K}^{-1}\right)^{k,\,l}_{\bm{n},\,\bm{n}{'}}
	\hat{P}^{(-)}_{k;\,\bm{n}{'}}+
	\hat{P}^{(-)}_{k;\,\bm{n}}
	\left(\tilde{K}^{-1}\right)^{k,\,l}_{\bm{n},\,\bm{n}{'}}
	\hat{P}^{(-)}_{l;\,\bm{n}{'}}\right)\right\}
	\vert0\rangle
\end{multline}
and
\begin{multline}
	\langle\{P^{k}\}\vert=
	\frac{1}{\det\left(\pi\tilde{K}\right)^{1/4}}
	\langle0\vert\\
	\times
	\exp\left\{-\sum_{\bm{n},\,\bm{n}{'}}\sum_{k,\,l=1}^{3}
	\left(
	\frac{1}{2}P_{k;\,\bm{n}}
	\left(\tilde{K}^{-1}\right)^{k,\,l}_{\bm{n},\,\bm{n}{'}}
	P_{k;\,\bm{n}{'}}-
	2P^{k}{}_{;\,\bm{n}}
	\left(\tilde{K}^{-1}\right)^{k,\,l}_{\bm{n},\,\bm{n}{'}}
	\hat{P}^{(+)}_{k;\,\bm{n}{'}}+
	\hat{P}^{(+)}_{k;\,\bm{n}}
	\left(\tilde{K}^{-1}\right)^{k,\,l}_{\bm{n},\,\bm{n}{'}}
	\hat{P}^{(+)}_{k;\,\bm{n}{'}}\right)\right\}
\end{multline}
as right and left eigenvectors of $\hat{P}_{k;\,\bm{n}}$ satisfying
\begin{equation}
	\hat{P}_{k;\,\bm{n}}\vert\{P^{k}\}\rangle=
	\vert\{P_{k}\}\rangle P_{k}{}_{;\,\bm{n}},\quad
	\langle\{P_{k}\}\vert\hat{P}_{k}{}_{;\,\bm{n}}=
	P_{k;\,\bm{n}}\langle\{P_{k}\}\vert.
\end{equation}

The resolution of unity on the Fock space of the spatial component can be expressed as
\begin{equation}
\label{eq:resPU}
	\int_{-\infty}^{\infty}\!\!\prod_{\bm{n}=-N}^{N}\prod_{k=1}^{3}
	dP_{k;\,\bm{n}}
	\vert\{P_{k}\}\rangle\langle\{P_{k}\}\vert=\bm{1}
\end{equation}
and the inner product between eigenvectors above reads
\begin{equation}
	\langle\{P_{k}\}\vert\{P{'}_{k}\}\rangle=
	\prod_{\bm{n}=-N}^{N}\prod_{k=1}^{3}\delta(P_{k;\,\bm{n}}-
	P{'}_{k;\,\bm{n}}).
\end{equation}
By making use of the commutation relation
\begin{equation}
	[\hat{U}^{(+)}_{k;\,\bm{n}},\,\hat{P}^{(-)}_{l;\,\bm{n}{'}}]=
	\frac{i}{2}W^{k,\,l}_{\bm{n},\,\bm{n}{'}},\quad
	W^{k,\,l}_{\bm{n},\,\bm{n}{'}}\equiv
	\sum_{\bm{r}=-N}^{N}\left(\delta_{k,\,l}+
	\frac{P_{k;\,\bm{r}}P^{*}_{l;\,\bm{r}}}{M^{2}}\right)
	F_{\bm{n}}^{\bm{r}}F_{\bm{n}{'}}^{\bm{r}{'}*},
\end{equation}
we can also calculate the inner product between eigenvectors of $\hat{U}_{k;\,\bm{n}}$ and those of $\hat{P}_{k;\,\bm{n}{'}}$ to find
\begin{equation}
\begin{aligned}
	\langle\{U_{k}\}\vert\{P_{k}\}\rangle=&
	\frac{1}{\sqrt{\det(2\pi W)}}
	\exp\left\{i\sum_{\bm{n,\,\bm{n}{'}}=-N}^{N}\sum_{k,\,l=1}^{3}
	U_{k;\,\bm{n}}
	\left(W^{-1}\right)^{k,\,l}_{\bm{n},\,\bm{n}{'}}
	P_{l;\,\bm{n}{'}}\right\},\\
	\langle\{P_{k}\}\vert\{U_{k}\}\rangle=&
	\frac{1}{\sqrt{\det(2\pi W)}}
	\exp\left\{-i\sum_{\bm{n,\,\bm{n}{'}}=-N}^{N}\sum_{k,\,l=1}^{3}
	U_{k;\,\bm{n}}
	\left(W^{-1}\right)^{k,\,l}_{\bm{n},\,\bm{n}{'}}
	P_{l;\,\bm{n}{'}}\right\}.
\end{aligned}
\end{equation}
Here we should add a comment that the factor $(W^{-1}P)_{k;\,\bm{n}}$ can be written as $\varPi_{k;\,\bm{n}}$ if we put
\begin{equation}
	P_{k;\,\bm{n}}=\left(
	\delta_{k,\,l}-\frac{\nabla_{k}\tilde{\nabla}_{l}}{M^{2}}\right)
	\varPi_{l;\,\bm{n}}
\end{equation}
and the Jacobian of this change of variables is equal to $\det W$.
On this observation, we may define
\begin{equation}
	\vert\{\varPi_{k}\}\rangle\equiv
	\sqrt{\det W}\vert\{P_{k}\}\rangle,\quad
	\langle\{\varPi_{k}\}\vert\equiv
	\sqrt{\det W}\langle\{P_{k}\}\vert.
\end{equation}
Then, by making use of the relation
\begin{equation}
	\hat{\varPi}_{k;\,\bm{n}}=\left(\delta_{k,\,l}+
	\frac{\nabla_{k}\tilde{\nabla}_{l}}{-\Laplace+M^{2}}\right)
	\hat{P}_{k;\,\bm{n}},
\end{equation}
we observe
\begin{equation}
	\hat{\varPi}_{k;\bm{n}}\vert\{\varPi_{k}\}\rangle=
	\vert\{\varPi_{k}\}\rangle\varPi_{k;\,\bm{n}},\quad
	\langle\{\varPi_{k}\}\vert\hat{\varPi}_{k;\bm{n}}=
	\varPi_{k;\,\bm{n}}\langle\{\varPi_{k}\}\vert,\quad
	\varPi_{k;\,\bm{n}}\equiv
	\left(\delta_{k,\,l}+
	\frac{\nabla_{k}\tilde{\nabla}_{l}}{-\Laplace+M^{2}}\right)
	P_{k;\,\bm{n}},
\end{equation}
where $P_{k;\,\bm{n}}$ being the eigenvalue of $\hat{P}_{k;\,\bm{n}}$ on $\vert\{P_{k}\}\rangle$.
It will be now straightforward to see that there holds
\begin{equation}
\label{eq:resPi0}
	\int_{-\infty}^{\infty}\!\!\prod_{\bm{n}=-N}^{N}\prod_{k=1}^{3}
	d\varPi_{k;\,\bm{n}}
	\langle\{U_{k}\}\vert\{\varPi_{k}\}\rangle
	\langle\{\varPi_{k}\}\vert\{U{'}_{k}\}\rangle=
	\prod_{\bm{n}=-N}^{N}\prod_{k=1}^{3}
	\delta(U_{k;\,\bm{n}}-U{'}_{k;\,\bm{n}})
\end{equation}
as well as
\begin{equation}
	\int_{-\infty}^{\infty}\!\!\prod_{\bm{n}=-N}^{N}\prod_{k=1}^{3}
	dU_{k;\,\bm{n}}
	\langle\{\varPi_{k}\}\vert\{U_{k}\}\rangle
	\langle\{U_{k}\}\vert\{\varPi{'}_{k}\}\rangle=
	\prod_{\bm{n}=-N}^{N}\prod_{k=1}^{3}
	\delta(\varPi_{k;\,\bm{n}}-\varPi{'}_{k;\,\bm{n}}).
\end{equation}

Leaving the consideration on the spatial components of the vector field, we now proceed to the construction of eigenvectors of $\hat{B}_{\bm{n}}$ and $\hat{P}_{B;\,\bm{n}}$. We again define positive and negative frequency parts of $\hat{B}_{\bm{n}}$ as
\begin{equation}
	\hat{B}^{(+)}_{\bm{n}}\equiv
	\sum_{\bm{r}=-N}^{N}
	\frac{M}{\sqrt{2a\omega{'}_{\bm{r}}}}
	\hat{b}_{\bm{r}}
	F_{\bm{n}}^{\bm{r}},\quad
	\hat{B}^{(-)}_{\bm{n}}\equiv
	\sum_{\bm{r}=-N}^{N}
	\frac{M}{\sqrt{2a\omega{'}_{\bm{r}}}}
	\hat{b}^{\dagger}{}_{\bm{r}}
	F_{\bm{n}}^{\bm{r}*}.
\end{equation}
Since their commutation relation is given by
\begin{equation}
	[\hat{B}^{(+)}_{\bm{n}},\,\hat{B}^{(-)}_{\bm{n}{'}}]=
	-\frac{1}{2}K^{-1}_{B;\,\bm{n},\,\bm{n}{'}},\quad
	K^{-1}_{B;\,\bm{n},\,\bm{n}{'}}\equiv
	\sum_{\bm{r}=-N}^{N}\frac{M^{2}}{a\omega{'}_{\bm{r}}}
	F_{\bm{n}}^{\bm{r}}F_{\bm{n}{'}}^{\bm{r}*},
\end{equation}
we have to adopt the technique for the negative norm case in the appendix.

We first define the coherent state
\begin{equation}
	\vert\{B^{(+)}\}\rangle=\exp\left\{-2\sum_{\bm{n},\,\bm{n}{'}=-N}^{N}
	\hat{B}^{(-)}{}_{\bm{n}}K_{B;\,\bm{n},\,\bm{n}{'}}
	B^{(+)}_{\bm{n}{'}}\right\}
	\vert0\rangle
\end{equation}
to satisfy
\begin{equation}
	\hat{B}^{(+)}_{\bm{n}}\vert\{B^{(+)}\}\rangle=
	\vert\{B^{(+)}\}\rangle B^{(+)}_{\bm{n}}.
\end{equation}
The conjugate of this coherent state is given by
\begin{equation}
	\langle\underline{\{B^{(-)}\}}\vert=\langle0\vert
	\exp\left\{2\sum_{\bm{n},\,\bm{n}{'}=-N}^{N}
	B^{(-)}_{\bm{n}}K_{B;\,\bm{n},\,\bm{n}{'}}
	\hat{B}^{(+)}_{\bm{n}{'}}\right\},\quad
	B^{(-)}_{\bm{n}}=\left(B^{(+)}_{\bm{n}}\right)^{*}.
\end{equation}
On this state $\hat{B}^{(-)}_{\bm{n}}$ takes the eigenvalue $-B^{(-)}_{\bm{n}}$;
\begin{equation}
	\langle\underline{\{B^{(-)}\}}\vert
	\hat{B}^{(-)}_{\bm{n}}=
	-B^{(-)}_{\bm{n}}\langle\underline{\{B^{(-)}\}}\vert.
\end{equation}
The inner product between these coherent state is given by
\begin{equation}
	\langle\underline{\{B^{(-)}\}}\vert\{B{'}^{(+)}\}\rangle=
	\exp\left\{2\sum_{\bm{n}=-N}^{N}
	B^{(-)}_{\bm{n}{'}}K_{B;\,\bm{n},\,\bm{n}{'}}
	B{'}^{(+)}_{\bm{n}{'}}\right\}.
\end{equation}

We now define
\begin{equation}
	\vert\{B\}\rangle\langle\underline{\{B\}}\vert\equiv
	\int\!\!\prod_{\bm{n}=-N}^{N}\frac{d\lambda_{\bm{n}}}{2\pi}
	\exp\left\{\sum_{\bm{n}=-N}^{N}\lambda_{\bm{n}}\left(
	\hat{B}_{\bm{n}}-iB_{\bm{n}}\right)\right\},
\end{equation}
where $B_{\bm{n}}$ is a real valued $c$-number function.
Decomposition of the operator above and Gaussian integrals over $\lambda_{\bm{n}}$s can be carried out to result in
\begin{multline}
\label{eq:projector03}
	\langle\underline{\{B^{(-)}\}}\vert\{B\}\rangle
	\langle\underline{\{B\}}\vert\vert\{B{'}^{(+)}\}\rangle
	=
	\frac{1}{\sqrt{\det\left(\pi K_{B}^{-1}\right)}}\\
	\times
	\exp\left\{-\sum_{\bm{n},\,\bm{n}{'}}
	\left\{B_{\bm{n}}-i\left(
	B^{(-)}_{\bm{n}}-B{'}{}^{(+)}_{\bm{n}}\right)\right\}
	K_{B;\,\bm{n},\,\bm{n}{'}}\left\{
	B_{\bm{n}}-i\left(
	B^{(-)}_{\bm{n}}-B{'}^{(+)}_{\bm{n}}\right)\right\}\right\}
	\langle\{B^{(-)}\}\vert\{B{'}^{(+)}\}\rangle,
\end{multline}
By remembering the inner product $\langle\underline{\{B^{(-)}\}}\vert\{B{'}^{(+)}\}\rangle$ and replacing eigenvalues $-B^{(-)}_{\bm{n}}$ and $B^{(+)}_{\bm{n}}$ with the corresponding operators, we find
\begin{equation}
	\vert\{B\}\rangle=
	\frac{1}{\det\left(\pi K_{B}^{-1}\right)^{1/4}}
	\exp\left\{-\sum_{\bm{n},\,\bm{n}{'}}\left(
	\frac{1}{2}B_{\bm{n}}K_{B;\,\bm{n},\,\bm{n}{'}}B_{\bm{n}{'}}+
	2iB_{\bm{n}}K_{B;\,\bm{n},\,\bm{n}{'}}\hat{B}^{(-)}_{\bm{n}{'}}-
	\hat{B}^{(-)}_{\bm{n}}K_{B;\,\bm{n},\,\bm{n}{'}}
	\hat{B}^{(-)}_{\bm{n}{'}}\right)\right\}
	\vert0\rangle
\end{equation}
and
\begin{equation}
	\langle\underline{\{B\}}\vert=
	\frac{1}{\det\left(\pi K_{B}^{-1}\right)^{1/4}}
	\langle0\vert
	\exp\left\{-\sum_{\bm{n},\,\bm{n}{'}}\left(
	\frac{1}{2}B_{\bm{n}}K_{B;\,\bm{n},\,\bm{n}{'}}B_{\bm{n}{'}}+
	2iB_{\bm{n}}K_{B;\,\bm{n},\,\bm{n}{'}}\hat{B}^{(+)}_{\bm{n}{'}}-
	\hat{B}^{(+)}_{\bm{n}}K_{B;\,\bm{n},\,\bm{n}{'}}
	\hat{B}^{(+)}_{\bm{n}{'}}\right)\right\}
\end{equation}
as solutions for
\begin{equation}
	\hat{B}_{\bm{n}}\vert\{B\}\rangle=
	\vert\{B\}\rangle iB_{\bm{n}},\quad
	\langle\underline{\{B\}}\vert\hat{B}_{\bm{n}}=
	iB_{\bm{n}}\langle\underline{\{B\}}\vert.
\end{equation}
Note that the eigenvalue of $\hat{B}_{\bm{n}}$ is given by an imaginary number $iB_{\bm{n}}$ although in the resolution of unity given below we integrate with respect to the real number $B_{\bm{n}}$.

Again through the Gaussian integrals over $B_{\bm{n}}$ in \eqref{eq:projector03}, we observe that there holds
\begin{equation}
\label{eq:resB}
	\int_{-\infty}^{\infty}\!\!\prod_{\bm{n}=-N}^{N}dB_{\bm{n}}
	\vert\{B\}\rangle\langle\underline{\{B\}}\vert=\bm{1}
\end{equation}
as the resolution of unity on the indefinite metric Fock space of the $B$ field. The inner product of the eigenvectors is given by
\begin{equation}
	\langle\underline{\{B\}}\vert\{B{'}\}\rangle=
	\prod_{\bm{n}=-N}^{N}\delta(B_{\bm{n}}-B{'}_{\bm{n}}).
\end{equation}

To obtain eigenvectors of $\hat{P}_{B;\,\bm{n}}$, we decompose this operator into positive and negative frequency parts. Then define the coherent state by taking into account of the negative metric of the Hilbert space.
The projection operator to be defined for this purpose is
\begin{equation}
	\vert\{P_{B}\}\rangle\langle\underline{\{P_{B}\}}\vert\equiv
	\int\!\!\prod_{\bm{n}=-N}^{N}\frac{d\lambda_{\bm{n}}}{2\pi}
	\exp\left\{\sum_{\bm{n}=-N}^{N}\lambda_{\bm{n}}\left(
	\hat{P}_{B;\,\bm{n}}+iP_{B;\,\bm{n}}\right)\right\},
\end{equation}
Through the similar process for eigenvectors of $\hat{B}_{\bm{n}}$ above, we can find
\begin{equation}
	\vert\{P_{B}\}\rangle=
	\frac{1}{\det\left(\pi\tilde{K}_{B}\right)^{1/4}}
	\exp\left\{-\sum_{\bm{n},\,\bm{n}{'}}\left(
	\frac{1}{2}P_{B;\,\bm{n}}\tilde{K}^{-1}_{B;\,\bm{n},\,\bm{n}{'}}
	P_{B;\,\bm{n}{'}}-
	2iP_{B;\,\bm{n}}\tilde{K}^{-1}_{B;\,\bm{n},\,\bm{n}{'}}
	\hat{P}^{(-)}_{B;\,\bm{n}{'}}-
	\hat{P}^{(-)}_{B;\,\bm{n}}\tilde{K}^{-1}_{B;\,\bm{n},\,\bm{n}{'}}
	\hat{P}^{(-)}_{B;\,\bm{n}{'}}\right)\right\}
	\vert0\rangle
\end{equation}
and
\begin{equation}
	\langle\underline{\{P_{B}\}}\vert=
	\frac{1}{\det\left(\pi\tilde{K}_{B}\right)^{1/4}}
	\langle0\vert
	\exp\left\{-\sum_{\bm{n},\,\bm{n}{'}}\left(
	\frac{1}{2}P_{B;\,\bm{n}}\tilde{K}^{-1}_{B;\,\bm{n},\,\bm{n}{'}}
	P_{B;\,\bm{n}{'}}-
	2iP_{B;\,\bm{n}}\tilde{K}^{-1}_{B;\,\bm{n},\,\bm{n}{'}}
	\hat{P}^{(+)}_{B;\,\bm{n}{'}}-
	\hat{P}^{(+)}_{B;\,\bm{n}}\tilde{K}^{-1}_{B;\,\bm{n},\,\bm{n}{'}}
	\hat{P}^{(+)}_{B;\,\bm{n}{'}}\right)\right\},
\end{equation}
where $\tilde{K}^{-1}_{B;\,\bm{n},\,\bm{n}{'}}$ arises from the commutation relation
\begin{equation}
	[\hat{P}^{(+)}_{B;\,\bm{n}},\,\hat{P}^{(-)}_{B;\,\bm{n}{'}}]=
	-\frac{1}{2}\tilde{K}^{-1}_{B;\,\bm{n},\,\bm{n}{'}},\quad
	\tilde{K}^{-1}_{B;\,\bm{n},\,\bm{n}{'}}\equiv
	\sum_{\bm{r}=-N}^{N}\frac{a\omega{'}_{\bm{r}}M^{2}}{2}
	F_{\bm{n}}^{\bm{r}}F_{\bm{n}{'}}^{\bm{r}*}.
\end{equation}
The bra and ket vectors above satisfy
\begin{equation}
	\hat{P}_{B;\,\bm{n}}\vert\{P_{B}\}\rangle=
	\vert\{P_{B}\}\rangle(-i)P_{B;\,\bm{n}},\quad
	\langle\underline{\{P_{B}\}}\vert\hat{P}_{B;\,\bm{n}}=
	-iP_{B;\,\bm{n}}\langle\underline{\{P_{B}\}}\vert.
\end{equation}
We thus find again that the eigenvalue of $\hat{P}_{B;\,\bm{n}}$ is pure imaginary.

The resolution of unity for this degree of freedom can be expressed in terms of these eigenvectors as
\begin{equation}
\label{eq:resPB}
	\int_{-\infty}^{\infty}\!\!\prod_{\bm{n}=-N}^{N}dP_{B;\,\bm{n}}
	\vert\{P_{B}\}\rangle\langle\underline{\{P_{B}\}}\vert=\bm{1}
\end{equation}
and the inner product of these eigenvectors is given by
\begin{equation}
	\langle\underline{\{P_{B}\}}\vert\{P{'}_{B}\}\rangle=
	\prod_{\bm{n}=-N}^{N}\delta(P_{B;\,\bm{n}}-P{'}_{B;\,\bm{n}}).
\end{equation}
By making use of the commutation relation
\begin{equation}
	[\hat{B}^{(+)}{}_{\bm{n}},\,\hat{P}^{(-)}_{B;\,\bm{n}{'}}]=
	-\frac{i}{2}M^{2}\delta_{\bm{n},\,\bm{n}{'}},
\end{equation}
we can obtain the inner products between eigenvectors of $\hat{B}_{\bm{n}}$ and those of $\hat{P}_{B;\,\bm{n}}$ as
\begin{equation}
\begin{aligned}
	\langle\underline{\{B\}}\vert\{P_{B}\}\rangle
	=&
	\frac{1}{\sqrt{\det(2\pi M^{2})}}
	\exp\left(-\frac{i}{M^{2}}\sum_{\bm{n}=-N}^{N}
	B_{\bm{n}}P_{B;\,\bm{n}}\right),\\
	\langle\underline{\{P_{B}\}}\vert\{B\}\rangle
	=&
	\frac{1}{\sqrt{\det(2\pi M^{2})}}
	\exp\left(\frac{i}{M^{2}}\sum_{\bm{n}=-N}^{N}
	B_{\bm{n}}P_{B;\,\bm{n}}\right),
\end{aligned}
\end{equation}
from which we are convinced again that there hold resolutions of unity shown above. By taking account of the relation $\hat{P}_{B;\,\bm{n}}=-M^{2}\hat{\varPi}_{B;\,\bm{n}}$, we may put $\varPi_{B;\,\bm{n}}\equiv-M^{-2}P_{B;\,\bm{n}}$ in the above expressions and define
\begin{equation}
	\vert\{\varPi_{B}\}\rangle\equiv\sqrt{\det(M^{2})}
	\vert\{P_{B}\}\rangle,\quad
	\langle\underline{\{\varPi_{B}\}}\vert\equiv\sqrt{\det(M^{2})}
	\langle\underline{\{P_{B}\}}\vert
\end{equation}
to find
\begin{equation}
	\hat{\varPi}_{B;\,\bm{n}}\vert\{\varPi_{B}\}\rangle=
	\vert\{\varPi_{B}\}\rangle(-i)\varPi_{B;\,\bm{n}},\quad
	\langle\underline{\{\varPi_{B}\}}\vert\hat{\varPi}_{B;\,\bm{n}}=
	-i\varPi_{B;\,\bm{n}}\langle\{\varPi_{B}\}\vert.
\end{equation}

We have thus obtained the desired eigenvectors of field operators to reflect the indefinite metric of the Hilbert space. Combining all results above, we can now express the resolution of unity on this Hilbert space. To simplify the notation, we will write $U_{k;\,\bm{n}}$ as $\varPhi_{k;\,\bm{n}}$ for $k=1,\,2,\,3$ and $B_{\bm{n}}$ will be expressed by $\varPhi_{B;\,\bm{n}}$ in the following. In this simplified style, we can write the combined form of resolutions of unity \eqref{eq:resU} and \eqref{eq:resB} as
\begin{equation}
\label{eq:resPhi}
	\int_{-\infty}^{\infty}\!\!
	\prod_{\bm{n}=-N}^{N}\prod_{I\in\{1,\,2,\,3,\,B\}}d\varPhi_{I;\,\bm{n}}
	\vert\{\varPhi\}\rangle
	\langle\underline{\{\varPhi\}}\vert=
	\bm{1}.
\end{equation}
In the same way, \eqref{eq:resPU} or \eqref{eq:resPi0} and \eqref{eq:resPB} can be unified to be written as
\begin{equation}
	\label{eq:resPi}
	\int_{-\infty}^{\infty}\!\!
	\prod_{\bm{n}=-N}^{N}\prod_{I\in\{1,\,2,\,3,\,B\}}d\varPi_{I;\,\bm{n}}
	\vert\{\varPi\}\rangle
	\langle\underline{\{\varPi\}}\vert=
	\bm{1}.
\end{equation}
The inner products is then rewritten as
\begin{equation}
	\langle\underline{\{\varPhi\}}\vert\{\varPi\}\rangle=
	\frac{1}{(2\pi)^{2(2N+1)^{3}}}
	\exp\left(i\sum_{\bm{n}=-N}^{N}\sum_{I}
	\varPi_{I;\,\bm{n}}\varPhi_{I;\,\bm{n}}\right),\quad
	\langle\underline{\{\varPi\}}\vert\{\varPhi\}\rangle=
	\langle\underline{\{\varPhi\}}\vert\{\varPi\}\rangle^{*}.
\end{equation}

We have thus completed the preparation of our tools for formulating path integral representation of the system described by the Hamiltonian \eqref{eq:ham01}.

\section{Euclidean path integrals of a massive vector field on a lattice}
In this section we consider the Euclidean path integrals of the system described by the Hamiltonian \eqref{eq:ham01}. In the following, we assume the length of imaginary time $\beta$ to be equal to $L$ and devide it into $2N+1$ equal length pieces so that $x_{4}=-ix_{0}=n_{4}a$ ($n_{4}=0,\,\pm1,\,\pm2,\,\dots,\pm N$) should define spatial hypersurface in four dimensional cubic lattice.
To be suitable for expressions within the time sliced path integral, we write $\phi_{\bm{n}}(n_{0}a)$ as $\phi_{n}$ and change our notation of $\nabla_{0}\phi_{\bm{n}}(x^{0})$ from the previous definition in \eqref{eq:nabla01} to the corresponding difference, after generalizing the difference in spatial coordinates to four dimensional one;  $\nabla_{\mu}\phi_{n}\equiv\phi_{n}-\phi_{n-\hat{\mu}}$, $\tilde{\nabla}_{\mu}\phi_{n}\equiv\phi_{n+\hat{\mu}}-\phi_{n}$ for $\mu=1,\,2,\,3,\,4$.
We first construct time sliced path integral by directly utilizing the resolutions of unity obtained in the previous section. We will then try to 
formulate the same path integral in terms of $A_{\mu;\,n}$ instead of $U_{\mu;\,n}$.

\subsection{Use of field diagonal representation}
To formulate path integral representation of the system, we rewrite the Hamiltonian in terms of Schr\"{o}dinger operators as
\begin{multline}
\label{eq:ham02}
	\hat{H}=\frac{1}{a}\left\{
	\frac{1}{2}\hat{\varPi}_{k;\,\bm{n}}
	\left(\delta_{k,\,l}-
	\frac{\nabla_{k}\tilde{\nabla}_{l}}{M^{2}}\right)\hat{\varPi}_{l;\,\bm{n}}
	+\frac{1}{2}\hat{\varPhi}_{k;\,\bm{n}}
	\left(\delta_{k,\,l}-
	\frac{\nabla_{k}\tilde{\nabla}_{l}}{-\Laplace+M^{2}}
	\right)_{\bm{n},\,\bm{n}{'}}
	(-\Laplace+M^{2})\hat{\varPhi}_{l;\,\bm{n}}\right.\\
	\left.
	-\frac{1}{2}M^{2}\hat{\varPi}_{B;\,\bm{n}}^{2}-
	\frac{1}{2M^{2}}\hat{\varPhi}_{B;\,\bm{n}}
	(-\Laplace+\alpha M^{2})\hat{\varPhi}_{B;\,\bm{n}}\right\}.
\end{multline}
We assume hereafter the rule of sum over repeated indicies including the index for lattice points. Since our concern is on the quantum theory of the original vector field $A_{\mu;n}$, we add source terms
\begin{equation}
	\frac{1}{a}\left(
	J_{\mu;\,\bm{n}}\hat{A}_{\mu;\,\bm{n}}+
	J_{B;\,\bm{n}}\hat{B}_{\bm{n}}\right)
\end{equation}
at each Euclidean time $x_{4}=n_{4}a$ to the Hamiltonian. Here $J_{0;\,\bm{n}}$ couples to $\hat{A}_{0;\,\bm{n}}$.
We have expressions for $\hat{A}_{\mu;\,n}$ in terms of $\hat{U}_{\mu;\,n}$ and $\hat{B}_{n}$ in addition to the relations
\begin{equation}
	\hat{U}_{0;\,n}=-\frac{1}{M^{2}}\tilde{\nabla}_{k}\hat{\varPi}_{k},\quad
	\nabla_{0}\hat{B}_{n}=-M^{2}\varPi_{B;\,n}.
\end{equation}
We can therefore express the source terms above as
\begin{equation}
	J_{\mu;\,\bm{n}}\hat{A}_{\mu;\,\bm{n}}+
	J_{B;\,\bm{n}}\hat{B}_{\bm{n}}=
	\frac{1}{M^{2}}\nabla_{k}J_{0;\,\bm{n}}
	\hat{\varPi}_{k;\,\bm{n}}-
	J_{0;\,\bm{n}}\hat{\varPi}_{B;\,\bm{n}}
	+J_{k;\,\bm{n}}\hat{\varPhi}_{k;\,\bm{n}}
	+\left(J_{B;\,\bm{n}}-
	\frac{1}{M^{2}}\tilde{\nabla}_{k}J_{k;\,\bm{n}}\right)
	\hat{\varPhi}_{B;\,\bm{n}}.
\end{equation}
Note that we have defined source terms in terms of Schr\"{o}dinger operators.

By taking the effect of the source terms into account, we first evaluate a short time kernel
\begin{equation}
\label{eq:kernel01}
	K^{(\mathrm{E})}_{J}[\{\varPhi\},\{\varPhi{'}\};a]\equiv
	\langle\underline{\{\varPhi\}}\vert\left(
	1-a\hat{H}-
	J_{\mu;\,\bm{n}}\hat{A}_{\mu;\,\bm{n}}-
	J_{B;\,\bm{n}}\hat{B}_{\bm{n}}\right)
	\vert\{\varPhi{'}\}\rangle,
\end{equation}
which can be expressed as
\begin{multline}
\label{eq:kernel02}
	K^{(\mathrm{E})}_{J}[\{\varPhi\},\{\varPhi{'}\};a]=
	\frac{1}{(2\pi)^{4(2N+1)^{3}}}
	\int_{-\infty}^{\infty}\!\!\prod_{\bm{n}=-N}^{N}
	\prod_{I\in\{1,\,2,\,3,\,B\}}
	d\varPi_{I;\,\bm{n}}
	\exp\left\{
	i\varPi_{I;\,\bm{n}}\nabla_{4}\varPhi_{I;\,\bm{n}}
	-\varPi_{k;\,\bm{n}}\frac{1}{M^{2}}\nabla_{k}J_{0;\bm{n}}
	-i\varPi_{B;\,\bm{n}}J_{0;\,\bm{n}}
	\right.\\
	-\frac{1}{2}\varPi_{k;\,\bm{n}}
	\left(
	\delta_{k,\,l}-\frac{\nabla_{k}\tilde{\nabla}_{l}}{M^{2}}\right)
	\varPi_{l;\,\bm{n}}-\frac{1}{2}\varPhi_{k;\,\bm{n}}\left(
	\delta_{k,\,l}+\frac{\nabla_{k}\tilde{\nabla}_{l}}{-\Laplace+M^{2}}\right)
	\left(-\Laplace+M^{2}\right)\varPhi_{l;\,\bm{n}}\\
	\left.
	-\frac{1}{2}M^{2}\varPi_{B;\,\bm{n}}^{2}
	-\frac{1}{2M^{2}}\varPhi_{B;\,\bm{n}}
	\left(-\Laplace+\alpha M^{2}\right)\varPhi_{B;\,\bm{n}}
	-J_{k;\,\bm{n}}\varPhi_{k;\,\bm{n}}
	-i\left(J_{B;\,\bm{n}}-
	\frac{1}{M^{2}}\tilde{\nabla}_{k}J_{k;\,\bm{n}}\right)\varPhi_{B;\,\bm{n}}
	\right\},
\end{multline}
where use has been made of the resolution of unity \eqref{eq:resPi} and $\nabla_{4}\varPhi_{I;\,\bm{n}}\equiv\varPhi_{I;\,\bm{n}}-\varPhi{'}_{I;\,\bm{n}}$. We identify here $\varPhi_{I;\,\bm{n}}$ above with $n_{4}$-th integration variables arises from multiple insertion of the resolution of unity \eqref{eq:resPhi} to assign $n_{4}$ as the fourth index of lattice points and write it as $\varPhi_{I;\,n}$ for simplicity.
The generating functional defined on the lattice corresponding to the source terms is written as
\begin{equation}
\label{eq:ZE}
	Z^{(\mathrm{E})}[J]\equiv\int_{-\infty}^{\infty}\!\!
	\prod_{n_{\mu}=-N}^{N}\prod_{I\in\{1,\,2,\,3,\,B\}}
	d\varPhi_{I;\,n}
	K^{(\mathrm{E})}_{J}[\{\varPhi_{N}\},\{\varPhi_{N-1}\};a]
	K^{(\mathrm{E})}_{J}[\{\varPhi_{N-1}\},\{\varPhi_{N-2}\};a]
	\cdots
	K^{(\mathrm{E})}_{J}[\{\varPhi_{-N}\},\{\varPhi_{N}\};a],
\end{equation}
where $\varPhi_{n_{4}}$ designates $\varPhi_{I;\,n}$s symbolically and we have adopted the periodic boundary condition for the time direction in addition to the spatial directions.
If we define row vectors $V^{\mathrm{T}}_{n}\equiv(\varPi_{1;\,n},\,\dots,\,\varPi_{B;\,n};\,\varPhi_{1;\,n},\,\dots,\,\varPhi_{B;\,n})$ and
\begin{equation}
	{\mathcal J}^{\mathrm{T}}\equiv\left(\frac{1}{M^{2}}\nabla_{1}J_{0;\,n},\,
	\frac{1}{M^{2}}\nabla_{2}J_{0;\,n},\,
	\frac{1}{M^{2}}\nabla_{3}J_{0;\,n},\,iJ_{0;\,n},\,J_{1;\,n},\,J_{2;\,n},\,
	J_{3;\,n},\,
	i\left(J_{B;\,n}-\frac{1}{M^{2}}\tilde{\nabla}_{k}J_{k;\,n}\right)\right)
\end{equation}
as well as an $8\times8$ matrix ${\mathcal M}^{-1}$ by
\begin{equation}
	{\mathcal M}^{-1}_{n,\,n{'}}\equiv
	\begin{pmatrix}
	W&0&-i\bm{1}\nabla_{4}&0\\
	0&M^{2}&0&-i\nabla_{4}\\
	i\bm{1}\tilde{\nabla}_{4}&0&\left(-\Laplace+M^{2}\right)W^{-1}&0\\
	0&i\tilde{\nabla}_{4}&0&\dfrac{1}{M^{2}}\left(-\Laplace+\alpha M^{2}\right)
	\end{pmatrix}_{\bm{n},\,\bm{n}{'}}\delta_{n_{4},\,n{'}_{4}},
\end{equation}
where $W$ and $W^{-1}$ are $3\times3$ matrices define in the previous section,
we can write the phase space path integral for the generating functional $Z^{(\mathrm{E})}[J]$ as
\begin{equation}
	Z^{(\mathrm{E})}[J]=\int_{-\infty}^{\infty}\!\!
	\prod_{n_{\mu}=-N}^{N}\prod_{I\in\{1,\,2,\,3,\,B\}}
	d\varPi_{I;\,n}d\varPhi_{I;\,n}
	e^{-{\mathcal A}^{(\mathrm{E})}_{J}},
\end{equation}
in which the phase space Euclidean action ${\mathcal A}^{(\mathrm{E})}_{J}$ being given by
\begin{equation}
	{\mathcal A}^{(\mathrm{E})}_{J}=\frac{1}{2}
	V^{\mathrm{T}}_{n}{\mathcal M}^{-1}_{n,\,n{'}}V_{n{'}}+{\mathcal J}^{\mathrm{T}}_{n}V_{n}.
\end{equation}
Upon finding the that ${\mathcal M}$ being given by
\begin{equation}
	{\mathcal M}_{n,\,n{'}}=\begin{pmatrix}
	\left(-\Laplace+M^{2}\right)W^{-1}\Delta^{(\mathrm{E})}
	&0&i\bm{1}\nabla_{4}\Delta^{(\mathrm{E})}&0\\
	0&
	\left(-\Laplace+\alpha M^{2}\right)
	\dfrac{1}{M^{2}}\Delta^{(\mathrm{E})}_{(\alpha)}
	&0&i\nabla_{4}\Delta^{(\mathrm{E})}_{(\alpha)}\\
	-i\bm{1}\tilde{\nabla}_{4}\Delta^{(\mathrm{E})}&0&
	W\Delta^{(\mathrm{E})}&0\\
	0&-i\tilde{\nabla}_{4}\Delta^{(\mathrm{E})}_{(\alpha)}&
	0&M^{2}\Delta^{(\mathrm{E})}_{(\alpha)}
	\end{pmatrix}_{n,\,n{'}},
\end{equation}
where we have defined $\Delta^{(\mathrm{E})}_{n,\,n{'}}$ and $\Delta^{(\mathrm{E})}_{(\alpha)\,n,\,n{'}}$, under the periodic boundary condition, by
\begin{equation}
	\left(-\DAlambert^{(\mathrm{E})}+M^{2}\right)
	\Delta^{(\mathrm{E})}_{n,\,n{'}}=\delta_{n,\,n{'}},\quad
	\left(-\DAlambert^{(\mathrm{E})}+\alpha M^{2}\right)
	\Delta^{(\mathrm{E})}_{(\alpha)\,n,\,n{'}}=\delta_{n,\,n{'}},\quad
	\DAlambert^{(\mathrm{E})}\equiv\tilde{\nabla}_{\mu}\nabla_{\mu},
\end{equation}
we can carry out the phase space path integral above to obtain
\begin{equation}
	Z^{(\mathrm{E})}[J]=Z^{(\mathrm{E})}_{0}e^{-W^{(\mathrm{E})}[J]},\quad
	Z^{(\mathrm{E})}_{0}\equiv\frac{1}
	{\sqrt{\det(-\DAlambert^{(\mathrm{E})}+M^{2})^{3}
	\det(-\DAlambert^{(\mathrm{E})}+\alpha M^{2})}},
\end{equation}
where the generating functional for connected Green's functions being given by
\begin{multline}
\label{eq:W}
	W^{(\mathrm{E})}[J]=\frac{1}{2}J_{0;\,n}\left\{
	\left(1-\frac{\nabla_{4}\tilde{\nabla}_{4}}{M^{2}}\right)
	\Delta^{(\mathrm{E})}_{n,\,n{'}}+
	\frac{\nabla_{4}\tilde{\nabla}_{4}}{M^{2}}
	\Delta^{(\mathrm{E})}_{(\alpha)\,n,\,n{'}}
	\right\}J_{0;\,n{'}}-
	\frac{1}{2}J_{k;\,n}\left\{
	\left(\delta_{k,\,l}-\frac{\nabla_{k}\tilde{\nabla}_{l}}{M^{2}}\right)
	\Delta^{(\mathrm{E})}_{n,\,n{'}}+
	\frac{\nabla_{k}\tilde{\nabla}_{l}}{M^{2}}
	\Delta^{(\mathrm{E})}_{(\alpha)\,n,\,n{'}}
	\right\}J_{l;\,n{'}}\\
	+iJ_{0;\,n}
	\frac{\nabla_{4}\tilde{\nabla}_{k}}{M^{2}}\left(
	\Delta^{(\mathrm{E})}_{n,\,n{'}}
	-\Delta^{(\mathrm{E})}_{(\alpha)\,n,\,n{'}}\right)
	J_{k;\,n{'}}+
	\frac{1}{2}J_{B;\,n}M^{2}
	\Delta^{(\mathrm{E})}_{(\alpha)\,n,\,n{'}}J_{B;\,n{'}}+
	iJ_{0;\,n}\nabla_{4}\Delta^{(\mathrm{E})}_{(\alpha)\,n,\,n{'}}J_{B;\,n{'}}+
	J_{k;\,n}\nabla_{k}\Delta^{(\mathrm{E})}_{(\alpha)\,n,\,n{'}}J_{B;\,n{'}}.
\end{multline}
Introducing $J^{(\mathrm{E})}_{\mu;\,n}$ by putting $J^{(\mathrm{E})}_{k;\,n}\equiv J_{k;\,n}$ and $J^{(\mathrm{E})}_{4;\,n}\equiv iJ_{0;\,n}$, we can simplify the right hand side above to write
\begin{multline}
\label{eq:WE}
	W^{(\mathrm{E})}[J]=
	-\frac{1}{2}J^{(\mathrm{E})}_{\mu;\,n}\left\{\left(
	\delta_{\mu,\,\nu}-\frac{1}{M^{2}}\nabla_{\mu}\tilde{\nabla}_{\nu}\right)
	\Delta^{(\mathrm{E})}_{n,\,n{'}}+
	\frac{1}{M^{2}}\nabla_{\mu}\tilde{\nabla}_{\nu}
	\Delta^{(\mathrm{E})}_{(\alpha)\,n,\,n{'}}\right\}
	J^{(\mathrm{E})}_{\nu;\,n{'}}\\
	+\frac{1}{2}J_{B;\,n}M^{2}
	\Delta^{(\mathrm{E})}_{(\alpha)\,n,\,n{'}}J_{B;\,n{'}}+
	J^{(\mathrm{E})}_{\mu;\,n}\nabla_{\mu}
	\Delta^{(\mathrm{E})}_{(\alpha)\,n,\,n{'}}J_{B;\,n{'}}.
\end{multline}

We may define Green's functions by
\begin{equation}
	\left\langle\mathrm{T}\left\{\hat{\varphi}_{I_{1};\,n_{1}}
	\hat{\varphi}_{I_{2};\,n_{2}}\cdots
	\hat{\varphi}_{I_{k};\,n_{k}}\right\}\right\rangle
	\equiv
	\frac{1}{Z^{(\mathrm{E})}_{0}}
	\left.
	(-1)^{k}\frac{\partial^{k}\ }{%
	\partial J_{I_{1};\,n_{1}}\partial J_{I_{2};\,n_{2}}\cdots
	\partial J_{I_{k};\,n_{k}}}
	\right\vert_{J=0}
	Z^{(\mathrm{E})}[J],
\end{equation}
in which $\hat{\varphi}_{I;\,n}$ designates $\hat{A}_{\mu;\,n}$ and $\hat{B}_{n}$ in a uniform manner. Corresponding to the definition of $J^{(\mathrm{E})}_{\mu;\,n}$ above, however, it will be convenient to define
$\hat{A}^{(\mathrm{E})}_{\mu;\,n}$ by putting $\hat{A}^{(\mathrm{E})}_{k;\,n}\equiv \hat{A}_{k;\,n}$ and $\hat{A}^{(\mathrm{E})}_{4;\,n}\equiv-i\hat{A}_{0;\,n}$ to find
\begin{equation}
	\left\langle\mathrm{T}\left\{
	\hat{A}^{(\mathrm{E})}_{\mu;\,n}
	\hat{A}^{(\mathrm{E})}_{\nu;\,n{'}}\right\}\right\rangle
	=
	\left(
	\delta_{\mu,\,\nu}-\frac{1}{M^{2}}
	\nabla_{\mu}\tilde{\nabla}_{\nu}
	\right)
	\Delta^{(\mathrm{E})}_{n,\,n{'}}+
	\frac{1}{M^{2}}
	\nabla_{\mu}\tilde{\nabla}_{\nu}
	\Delta^{(\mathrm{E})}_{(\alpha)\,n,\,n{'}}
\end{equation}
in addition to
\begin{equation}
	\left\langle\mathrm{T}\left\{
	\hat{B}_{n}\hat{B}_{n{'}}\right\}\right\rangle
	=
	-M^{2}\Delta^{(\mathrm{E})}_{(\alpha)\,n,\,n{'}},\quad
	\left\langle\mathrm{T}\left\{
	\hat{A}^{(\mathrm{E})}_{\mu;\,n}
	\hat{B}_{n{'}}\right\}\right\rangle
	=
	-\nabla_{\mu}\Delta^{(\mathrm{E})}_{(\alpha)\,n,\,n{'}}.
\end{equation}
Covariance in Green's functions above will be translated into the one for Green's functions in the Minkowski space by putting $x_{4}=ix_{0}$ as well as $\hat{A}^{(\mathrm{E})}_{4;\,n}=-i\hat{A}_{0;\,n}$ at the same time.

Covariance in Minkowski space will be seen more clearly by considering the effective action of the system under consideration.
To this end, we go back to the generating functional $W^{(\mathrm{E})}[J]$ in \eqref{eq:WE} and define classical solutions under the influence of external sources by
\begin{equation}
	\varphi_{\mathrm{cl.}\,I;\,n}^{(\mathrm{E})}\equiv
	\frac{\partial W[J]^{(\mathrm{E})}}{\partial J^{(\mathrm{E})}_{I;\,n}}.
\end{equation}
We then define the effective action through the Legendre transform by
\begin{equation}
	\varGamma^{(\mathrm{E})}[\varphi_{\mathrm{cl.}}^{(\mathrm{E})}]\equiv
	W^{(\mathrm{E})}[J]-
	J^{(\mathrm{E})}_{I;\,n}\varphi_{\mathrm{cl.}\,I;\,n}^{(\mathrm{E})}.
\end{equation}
Equations of motion for the classical solutions read
\begin{equation}
\begin{aligned}
	A^{(\mathrm{E})}_{\mathrm{cl.}\,\mu;\,n}=&-
	\left\{\left(
	\delta_{\mu,\,\nu}-\frac{1}{M^{2}}\nabla_{\mu}\tilde{\nabla}_{\nu}\right)
	\Delta^{(\mathrm{E})}_{n,\,n{'}}+
	\frac{1}{M^{2}}\nabla_{\mu}\tilde{\nabla}_{\nu}
	\Delta^{(\mathrm{E})}_{(\alpha)\,n,\,n{'}}\right\}
	J^{(\mathrm{E})}_{\nu;\,n{'}}+
	\nabla_{\mu}
	\Delta^{(\mathrm{E})}_{(\alpha)\,n,\,n{'}}J_{B;\,n{'}}\\
	B^{(\mathrm{E})}_{\mathrm{cl.}\,n}=&
	M^{2}\Delta^{(\mathrm{E})}_{(\alpha)\,n,\,n{'}}J_{B;\,n{'}}-
	\tilde{\nabla}_{\mu}\Delta^{(\mathrm{E})}_{(\alpha)\,n,\,n{'}}
	J^{(\mathrm{E})}_{\mu;\,n{'}}
\end{aligned}
\end{equation}
and can be inverted as
\begin{equation}
	J^{(\mathrm{E})}_{\mu;\,n}=-\left\{
	\delta_{\mu,\,\nu}\left(-\DAlambert^{(\mathrm{E})}+M^{2}\right)
	+\nabla_{\mu}\tilde{\nabla}_{\nu}\right\}
	U^{(\mathrm{E})}_{\mathrm{cl.}\,\nu;\,n},\quad
	J_{B;\,n}=\frac{1}{M^{2}}\left(-\DAlambert^{(\mathrm{E})}+M^{2}\right)
	B^{(\mathrm{E})}_{\mathrm{cl.}\,n}-
	\tilde{\nabla}_{\mu}U^{(\mathrm{E})}_{\mathrm{cl.}\,\mu;\,n},
\end{equation}
where we have defined
\begin{equation}
	U^{(\mathrm{E})}_{\mathrm{cl.}\,\mu;\,n}\equiv
	A^{(\mathrm{E})}_{\mathrm{cl.}\,\mu;\,n}-
	\frac{1}{M^{2}}\nabla_{\mu}B^{(\mathrm{E})}_{\mathrm{cl.}\,n}
\end{equation}
and use has been made of a relation
\begin{equation}
	\tilde{\nabla}_{\mu}J^{(\mathrm{E})}_{\mu;\,n}=-M^{2}
	\tilde{\nabla}_{\mu}A^{(\mathrm{E})}_{\mu;\,n}.
\end{equation}
We are now able to write the effective action in terms of the Euclidean vector field $A^{(\mathrm{E})}_{\mu;\,n}$ as well as $B^{(\mathrm{E})}_{n}$ to be
\begin{equation}
	\varGamma^{(\mathrm{E})}[A^{(\mathrm{E})}_{\mathrm{cl.}\,\mu},\,
	B^{(\mathrm{E})}_{\mathrm{cl.}}]
	=\frac{1}{2}U^{(\mathrm{E})}_{\mathrm{cl.}\,\mu}\left\{
	\delta_{\mu,\,\nu}\left(-\DAlambert^{(\mathrm{E})}+M^{2}\right)+
	\nabla_{\mu}\tilde{\nabla}_{\nu}\right\}U^{(\mathrm{E})}_{\nu;\,n}-
	\frac{1}{2M^{2}}B^{(\mathrm{E})}_{\mathrm{cl.}\,n}
	\left(-\DAlambert^{(\mathrm{E})}+\alpha M^{2}\right)
	B^{(\mathrm{E})}_{\mathrm{cl.}\,n}
\end{equation}

In order to see that the effective action obtained above reproduces the covariant action in the Minkowski space, let us now take the continuum limit by putting $a\to0$ while keeping $L$ intact. To do so, we write $A^{(\mathrm{E})}_{\mathrm{cl.}\,\mu;\,n}=aA^{(\mathrm{E})}_{\mathrm{cl.}\,\mu}(x)$ and $B^{(\mathrm{E})}_{\mathrm{cl.}\,B;\,n}=a^{2}B^{(\mathrm{E})}_{\mathrm{cl.}}(x)$ and afterwise we put $L\to\infty$ to obtain
\begin{multline}
\label{eq:GammaE}
	\varGamma^{(\mathrm{E})}[A^{(\mathrm{E})}_{\mathrm{cl.}\,\mu},\,
	B^{(\mathrm{E})}_{\mathrm{cl.}}]
	=
	\int_{-\infty}^{\infty}\!\!d^4x^{\ }_{\mathrm{E}}\left\{
	\frac{1}{2}U^{\mathrm{(E)}}_{\mathrm{cl.}\,\mu}(x)\left\{
	\delta_{\mu,\,\nu}\left(-\partial^{2}+m^{2}\right)+
	\partial_{\mu}\partial_{\nu}\right\}
	U^{(\mathrm{E})}_{\mathrm{cl.}\,\nu}(x)\right.\\
	\left.-\frac{1}{2m^{2}}
	B^{(\mathrm{E})}_{\mathrm{cl.}}(x)
	\left(-\partial^{2}+\alpha m^{2}\right)
	B^{(\mathrm{E})}_{\mathrm{cl.}}(x)\right\},
\end{multline}
where we have written $\partial^{2}=\partial_{\mu}\partial_{\mu}$ and we have put
\begin{equation}
	U^{\mathrm{(E)}}_{\mathrm{cl.}\,\mu}(x)\equiv
	A^{(\mathrm{E})}_{\mathrm{cl.}\,\mu}(x)-
	\frac{1}{m^{2}}\partial_{\mu}
	B^{(\mathrm{E})}_{\mathrm{cl.}}(x).
\end{equation}
Upon the transition to the Minkowski space, according to the change $x_{4}\to ix_{0}$, $\partial_{4}$ changes to $-i\partial_{0}$ and hence we define $A_{\mathrm{cl.}\,\mu}(x)$ and $B_{\mathrm{cl.}}(x)$ for Minkowski space by $A_{\mathrm{cl.}\,k}(x)\equiv A^{(\mathrm{E})}_{\mathrm{cl.}\,k}(x)\Big\vert_{x_{4}=ix_{0}}$ and $A_{\mathrm{cl.}\,0}(x)\equiv iA^{(\mathrm{E})}_{\mathrm{cl.}\,4}(x)\Big\vert_{x_{4}=ix_{0}}$ as well as $B_{\mathrm{cl.}\,\mu}(x)\equiv B^{(\mathrm{E})}_{\mathrm{cl.}}(x)\Big\vert_{x_{4}=ix_{0}}$, respectively, to yield
\begin{equation}
	U_{\mathrm{cl.}\,\mu}(x)\equiv
	A_{\mathrm{cl.}\,\mu}(x)-
	\frac{1}{m^{2}}\partial_{\mu}
	B_{\mathrm{cl.}}(x),
\end{equation}
in addition to rewriting $-\partial^{2}\to\partial_{\mu}\partial^{\mu}$ with the definition of derivative with respect to covariant components $\partial^{\mu}=\eta^{\mu,\,\nu}\partial_{\nu}$.
We thus observe that $\varGamma^{(\mathrm{E})}[\varphi^{(\mathrm{E})}_{\mathrm{cl.}}]$ above changes into
\begin{equation}
	\varGamma^{(\mathrm{E})}
	[A^{(\mathrm{E})}_{\mathrm{cl.}\,\mu},\,
	B^{(\mathrm{E})}_{\mathrm{cl.}}]\Big\vert_{x_{4}=ix_{0}}
	=
	-i\varGamma[A_{\mathrm{cl.}\,\mu},\,
	B_{\mathrm{cl.}}],
\end{equation}
where
\begin{equation}
	\varGamma[A_{\mathrm{cl.}\,\mu},\,B_{\mathrm{cl.}}]
	=
	\int_{-\infty}^{\infty}\!\!d^4x\left\{-\frac{1}{2}
	U_{\mathrm{cl.}\,\mu}(x)\left\{-\eta^{\mu,\,\nu}
	\left(\partial_{\rho}\partial^{\rho}+m^{2}\right)+
	\partial^{\mu}\partial^{\nu}\right\}
	U_{\mathrm{cl.}\,\nu}(x)+\frac{1}{2m^{2}}
	B_{\mathrm{cl.}}(x)
	\left(\partial_{\rho}\partial^{\rho}+\alpha m^{2}\right)
	B_{\mathrm{cl.}}(x)\right\}
\end{equation}
being the effective action of the present system in the Minkowski space.
We may define $G^{\mathrm{cl.}}_{\mu\,\nu}(x)\equiv\partial_{\mu}U_{\mathrm{cl.}\,\nu}(x)-\partial_{\nu}U_{\mathrm{cl.}\,\mu}(x)$ as well as $F^{\mathrm{cl.}}_{\mu\,\nu}(x)\equiv\partial_{\mu}A_{\mathrm{cl.}\,\nu}(x)-\partial_{\nu}A_{\mathrm{cl.}\,\mu}(x)$ to rewrite the effective action above as
\begin{equation}
\begin{aligned}
	\varGamma[A_{\mathrm{cl.}\,\mu},\,B_{\mathrm{cl.}}]
	=&
	\int_{-\infty}^{\infty}\!\!d^4x\left\{-\frac{1}{4}
	G^{\mathrm{cl.}\,\mu\,\nu}(x)
	G^{\mathrm{cl.}}_{\mu\,\nu}(x)+\frac{1}{2}m^{2}
	U_{\mathrm{cl.}\,\mu}(x)
	U_{\mathrm{cl.}}^{\mu}(x)-\frac{1}{2m^{2}}
	\partial_{\mu}B_{\mathrm{cl.}}(x)
	\partial^{\mu}B_{\mathrm{cl.}}(x)+\frac{\alpha}{2}
	B^{2}_{\mathrm{cl.}}(x)\right\}\\
	=&
	\int_{-\infty}^{\infty}\!\!d^4x\left\{-\frac{1}{4}
	F^{\mathrm{cl.}\,\mu\,\nu}(x)
	F^{\mathrm{cl.}}_{\mu\,\nu}(x)+\frac{1}{2}m^{2}
	A_{\mathrm{cl.}\,\mu}(x)
	A_{\mathrm{cl.}}^{\mu}(x)+
	B_{\mathrm{cl.}}(x)
	\partial_{\mu}A_{\mathrm{cl.}}^{\mu}(x)
	+\frac{\alpha}{2}
	B^{2}_{\mathrm{cl.}}(x)\right\}.
\end{aligned}
\end{equation}
This is nothing but the continuum version of the action that corresponds to the original Lagrangian \eqref{eq:lag01}.
We are therefore convinced that the effective action defined by the Euclidean path integral formulated upon the indefinite metric Hilbert space can correctly reproduce the covariant effective action for the Minkowski space.

It must be commented here that the covariance both in the generating functional $W^{(\mathrm{E})}[J]$ and the effective action $\varGamma^{(\mathrm{E})}[A^{(\mathrm{E})}_{\mathrm{cl.}\,\mu},\,
B^{(\mathrm{E})}_{\mathrm{cl.}}]$ in Euclidean space are entirely fake because we have expressed them in terms of $J^{(\mathrm{E})}_{\mu;\,n}$ and $A^{(\mathrm{E})}_{\mathrm{cl.}\,\mu;\,n}$, respectively. If we write them in terms of the original sources $J_{\mu;\,n}$ and the field $A_{\mathrm{cl.}\,\mu;\,n}\equiv\partial W^{(\mathrm{E})}[J]/\partial J_{\mu;\,n}$, they no longer possess the covariance suitable for Euclidean space. Nevertheless, when expressed in terms of these original variables, they recover the covariance in Minkowski space just by putting $x_{4}=ix_{0}$. Although we have carried out path integral with respect to phase space variables at once above, we could have integrated $\varPi_{I;\,n}$s first and then carried out the remaining path integral over the configuration space variables to find that the action in the exponent of the path integral does not possess covariance on the Euclidean space.
It should be noted here, however, the action in the exponent of a path integral is merely a part of the integrand and furthermore the path integral is the definite integral whose covariance can be judged by its result not by its integrand. We can therefore know the covariance of the path integral only by the effective action or by Green's functions defined by the path integral under consideration.

\subsection{Euclidean path integral in terms of $A_{\mu}$}
So far we have formulated the path integral by utilizing eigenvectors of $\hat{U}_{k;\,\bm{n}}$, $\hat{B}_{\bm{n}}$ and those of $\hat{\varPi}_{k;\,\bm{n}}$, $\hat{\varPi}_{B\,\bm{n}}$. Since $\hat{U}_{k;\,\bm{n}}$ is related with $\hat{A}_{k;\,\bm{n}}$ and $\hat{B}_{\bm{n}}$ via $\hat{U}_{k;\,\bm{n}}=\hat{A}_{k;\,\bm{n}}-\nabla_{k}B_{\bm{n}}/M^{2}$, we may regard $\vert\{\varPhi\}\rangle$ as an eigenvector of $\hat{A}_{k;\,\bm{n}}$ though the eigenvalue becomes a complex number. In the same way, we may define $\hat{A}_{0;\,\bm{n}}$ by setting $\hat{A}_{0;\,\bm{n}}\equiv-\hat{\varPi}_{B;\,\bm{n}}-\tilde{\nabla}_{k}\hat{\varPi}_{k;\,\bm{n}}/M^{2}$ to possess $\vert\{\varPi\}\rangle$ as an eigenvector with a complex eigenvalue. We may therefore expect that there exists an expression of the path integral considered in the previous subsection to be expressed in terms of $A_{\mu;\,\bm{n}}$s, i.e. eigenvalues of $\hat{A}_{\mu;\,\bm{n}}$s. To clarify this idea, we return to the definition of the generating functional $Z^{(\mathrm{E})}[J]$ in \eqref{eq:ZE}.
Then, on expectations above and writing $\varPhi_{B;\,n}$ as $B_{n}$, we define
\begin{equation}
\label{eq:change}
	A^{(\mathrm{E})}_{k;\,n}\equiv
	\varPhi_{k;\,n}+\frac{i}{M^{2}}\nabla_{k}B_{n},\quad
	A^{(\mathrm{E})}_{4;\,n}\equiv\varPi_{B;\,n}+
	\frac{i}{M^{2}}\tilde{\nabla}_{k}\varPi_{k;\,n}
\end{equation}
to rewrite the phase space Euclidean action ${\mathcal A}^{(\mathrm{E})}$ as
\begin{multline}
	{\mathcal A}^{(\mathrm{E})}=\frac{1}{2}\left\{
	\varPi_{k;\,n}-i\left(
	\nabla_{4}A^{(\mathrm{E})}_{k;\,n}-\nabla_{k}A^{(\mathrm{E})}_{4;\,n}\right)
	\right\}^{2}+
	\frac{\alpha}{2}\left\{
	B_{n}+\frac{i}{\alpha}\left(\tilde{\nabla}_{\mu}A^{(\mathrm{E})}_{\mu;\,n}+
	J_{B;\,n}\right)\right\}^{2}\\
	+\frac{1}{2}\left(\nabla_{4}A^{(\mathrm{E})}_{k;\,n}-
	\nabla_{k}A^{(\mathrm{E})}_{4;\,n}\right)^{2}+
	\frac{1}{2\alpha}
	\left(\tilde{\nabla}_{\mu}A^{(\mathrm{E})}_{\mu;\,n}+J_{B;\,n}\right)^{2}\\
	+\frac{1}{2}A^{(\mathrm{E})}_{k;\,n}\left\{
	\delta_{k,\,l}\left(-\Laplace+M^{2}\right)+\nabla_{k}\tilde{\nabla}_{l}
	\right\}A^{(\mathrm{E})}_{l;\,n}
	+\frac{1}{2}M^{2}A^{(\mathrm{E})\,2}_{4;\,n}+
	J^{(\mathrm{E})}_{\mu;\,n}A^{(\mathrm{E})}_{\mu;\,n},
\end{multline}
in which we have defined $J^{(\mathrm{E})}_{\mu;\,n}$ just in the same way as before.

Here a comment is in need; the action above should be compared with the one given by Eq.~(17) of Ref.~\cite{Kashiwa:81} by Kashiwa and also with Eq.~($4\cdot11$) of Ref.~\cite{Kashiwa-Sakamoto} by Kashiwa and Sakamoto. In the action ${\mathcal A}^{(\mathrm{E})}$ above, $A^{(\mathrm{E})}_{4;\,n}$ plays precisely the same role as the one played by $A=B_{4}$ in Eq.~(17) of Ref.~\cite{Kashiwa:81} and $A_{4;\,\bm{n}}(j)=(\nabla^{3})^{-1}C_{\bm{n}}(j)$ in Eq.~($4\cdot11$) of Ref.~\cite{Kashiwa-Sakamoto}. On the basis that they formulated their path integrals on the positive metric Hilbert space upon which physical degrees are represented, they asserted that only the positive definite Hilbert spaces were needed in the formulation of their path integrals. It is, however, quite clear now that the dummy variables introduced via Gaussian identities should be identified with $A^{(\mathrm{E})}_{4;\,\bm{n}}$, which gives the eigenvalue of $\hat{A}_{0;\,\bm{n}}$ as $iA^{(\mathrm{E})}_{4;\,\bm{n}}$, in our formulation. We therefore understand that they have cleverly invented a method of introducing degrees with indefinite metric without the explicit use of indefinite metric Hilbert space and hence that their results are in this sense equivalent to that of Ref.~\cite{Arisue:81} by Arisue {\em et al.\/}

Integrations with respect to $\varPi_{k;\,n}$s and $B_{n}$s will be carried out to yield the determinant factor $\{\sqrt{(2\pi)^{4}/\alpha}\}^{(2N+1)^{4}}$ and the remaining exponent as the integrand for integrals over $A^{(\mathrm{E})}_{\mu;\,n}$s. By writing the remaining exponent as $-S^{(\mathrm{E})}_{J}[A^{(\mathrm{E})}_{\mu}]$, we observe that $S^{(\mathrm{E})}_{J}[A^{(\mathrm{E})}_{\mu}]$ is expressed as
\begin{equation}
	S^{(\mathrm{E})}_{J}[A^{(\mathrm{E})}_{\mu}]=\frac{1}{2}
	A^{(\mathrm{E})}_{\mu;\,n}
	\left(D^{-1}_{(\mathrm{E})}\right)^{\mu,\,\nu}_{n,\,n{'}}
	A^{(\mathrm{E})}_{\nu;\,n{'}}+
	\tilde{J}^{(\mathrm{E})}_{\mu;\,n}A^{(\mathrm{E})}_{\mu;\,n}+
	\frac{1}{2\alpha}J_{B;\,n}^{2},
\end{equation}
where modified sources are given by
\begin{equation}
	\tilde{J}^{(\mathrm{E})}_{\mu;\,n}\equiv
	J^{(\mathrm{E})}_{\mu;\,n}-\frac{1}{\alpha}\nabla_{\mu}J_{B;\,n},
\end{equation}
while $D^{-1}_{(\mathrm{E})}$ being given by
\begin{equation}
	\left(D^{-1}_{(\mathrm{E})}\right)_{\mu,\,\nu}^{n,\,n{'}}\equiv
	\left\{\left(-\DAlambert^{(\mathrm{E})}+M^{2}\right)\delta_{\mu,\,\nu}
	-\frac{1-\alpha}{\alpha}
	\nabla_{\mu}\tilde{\nabla}_{\nu}\right\}\delta_{n,\,n{'}}.
\end{equation}
Thus the covariance of the Euclidean action is evident and we observe that the Euclidean path integral in terms of $A^{(\mathrm{E})}_{\mu;\,n}$ is nothing but the one of the Euclidean vector field usually obtained in the literature by the prescription $A_{0}=iA_{4}$.

As we will confirm at the end of this section, however, $A^{(\mathrm{E})}_{\mu;\,n}$ arises in our formulation from ($-i$ times) the eigenvalue of $\hat{A}_{0;\,\bm{n}}$ defined on an indefinite metric Hilbert space. Hence the Wick rotation has nothing to do with going to the Euclidean formulation of the path integral for the vector field. In this regard, we point it out that the Euclidean path integral is broadly regarded to be obtained by the Wick rotation at the same time we put $x^{0}=ix^{4}$ in the manifestly covariant path integral in Minkowski space. As we will see below the manifestly covariant path integral in Minkowski space does not satisfy the Euclidicity postulate\cite{Abers-Lee,Coleman:85} and hence cannot be converted into the Euclidean one just by putting $x^{0}=ix^{4}$. This is the reason why we need the Wick rotation in addition. In our formulation of the Euclidean path integral, the situation is ultimately different from path integrals obtained by such a prescription; $iA^{(\mathrm{E})}_{4;\,n}$ naturally arises as an eigenvalue of $\hat{A}_{0;\,\bm{n}}$ by the use of indefinite metric representation of commutation relations. Moreover, if the vector field under consideration is coupled to a Fermion in a minimal way, $J_{\mu;\,n}$ will consist of the corresponding current in addition to the external source. In accordance with the definition of $J^{(\mathrm{E})}_{\mu;\,n}$ above, we may define $\gamma^{4}\equiv i\gamma^{0}$ to convert the covariant derivative in the Fermion action suitable for the Euclidean path integral. We should therefore reinterpret the Euclidean path integral considered to be obtained by the Wick rotation from the viewpoint of the indefinite metric representation. As for the lattice gauge theory, the action of the Euclidean path integral coincides with the action of the Euclidean vector field above in the perturbative region. Hence it defines the path integral that fulfills the Euclidicity postulate\cite{Abers-Lee,Coleman:85} in a correct manner though the existence of the indefinite metric representation of commutation relations is not evident.

It is straightforward to find that $D_{(\mathrm{E})}$ is given by
\begin{equation}
	D^{\mu,\,\nu}_{(\mathrm{E})\,n,\,n{'}}=
	\left(\delta_{\mu,\,\nu}-\frac{\nabla_{\mu}\tilde{\nabla}_{\nu}}{M^{2}}
	\right)\Delta^{(\mathrm{E})}_{n,\,n{'}}+
	\frac{\nabla_{\mu}\tilde{\nabla}_{\nu}}{M^{2}}
	\Delta^{(\mathrm{E})}_{(\alpha)\,n,\,n{'}}
\end{equation}
and further that the generating functional $\tilde{W}^{(\mathrm{E})}[J]$ for the present case reads
\begin{equation}
\label{eq:w01}
	\tilde{W}^{(\mathrm{E})}[J]\equiv-\frac{1}{2}
	\tilde{J}^{(\mathrm{E})}_{\mu;\,n}D^{\mu\,\nu}_{(\mathrm{E})\,n,\,n{'}}
	\tilde{J}^{(\mathrm{E})}_{\nu;\,n{'}}+
	\frac{1}{2\alpha}J_{B;\,n}^{2}.
\end{equation}
Hence we again, by carrying out Gaussian integrals with respect to $A^{(\mathrm{E})}_{\mu;\,n}$s, obtain
\begin{equation}
	Z^{(\mathrm{E})}[J]=
	\frac{1}{\sqrt{\det(\alpha D_{(\mathrm{E})}^{-1})}}
	e^{-\tilde{W}^{(\mathrm{E})}[J]},
\end{equation}
where
\begin{equation}
	\det(\alpha D_{(\mathrm{E})}^{-1})=
	\det(-\DAlambert^{(\mathrm{E})}+M^{2})^{3}
	\det(-\DAlambert^{(\mathrm{E})}+\alpha M^{2}).
\end{equation}

The calculation of the effective action from $\tilde{W}^{(\mathrm{E})}[J]$ above goes in the similar manner as before and we obtain
\begin{equation}
	\tilde{\varGamma}^{(\mathrm{E})}[A^{(\mathrm{E})}_{\mathrm{cl.}\,\mu},\,
	B^{(\mathrm{E})}_{\mathrm{cl.}}]=
	\frac{1}{2}A^{(\mathrm{E})}_{\mathrm{cl.}\,\mu;\,n}\left\{
	\delta_{\mu,\nu}\left(-\DAlambert^{(\mathrm{E})}+M^{2}\right)+
	\nabla_{\mu}\tilde{\nabla}_{\nu}\right\}
	A^{(\mathrm{E})}_{\mathrm{cl.}\,\mu;\,n}+
	B^{(\mathrm{E})}_{\mathrm{cl.}\,n}\tilde{\nabla}_{\mu}
	A^{(\mathrm{E})}_{\mathrm{cl.}\,\mu;\,n}-
	\frac{\alpha}{2}B^{(\mathrm{E})\,2}_{\mathrm{cl.}\,n}
\end{equation}
as the effective action of the system.
It will be evident that the result obtained here is completely equivalent to the one we obtained in the previous subsection and hence $\tilde{\varGamma}^{(\mathrm{E})}[A^{(\mathrm{E})}_{\mathrm{cl.}\,\mu},\,B^{(\mathrm{E})}_{\mathrm{cl.}}]=\varGamma^{(\mathrm{E})}[A^{(\mathrm{E})}_{\mathrm{cl.}\,\mu},\,B^{(\mathrm{E})}_{\mathrm{cl.}}]$. Again, the covariance in the effective action above is, however, fake because $A^{(\mathrm{E})}_{\mathrm{cl.}\,\mu;\,n}$ is defined as the responce to $J^{(\mathrm{E})}_{\mu;\,n}$ not to $J_{\mu;\,n}$. Therefore the action of the Euclidean path integral has a covariant form by choosing suitable integration variables while the effective action resulting in a non-covariant form when expressed in terms of the real dynamical variables.

Before closing this section, we have to add comments on eigenvalues of $\hat{A}_{\mu;\,\bm{n}}$s;
note that, in carrying out Gaussian path integral 
\begin{equation}
	\frac{1}{\sqrt{\det\{(2\pi)^{4}\alpha\}}}
	\int_{-\infty}^{\infty}\!\!\prod_{n_{\mu}=-N}^{N}\prod_{\mu=1}^{4}
	dA^{(\mathrm{E})}_{\mu;\,n}
	e^{-S^{(\mathrm{E})}_{J}[A^{(\mathrm{E})}_{\mu}]},
\end{equation}
we have treated as if $A^{(\mathrm{E})}_{\mu;\,n}$s take real values from $-\infty$ to $+\infty$ despite the existence of the imaginary parts in their definition. Shifts in the imaginary direction in \eqref{eq:change} can be, just like an imaginary shift in a Gaussian integral, safely discarded as far as they remain to be finite. In this regard, we should observe here that the eigenvalue of $\hat{A}_{0;\bm{n}}$ should be regarded as pure imaginary; from the definition of $\hat{A}_{0;\,\bm{n}}$, given by
\begin{equation}
	\hat{A}_{0;\,\bm{n}}=-\hat{\varPi}_{B;\,\bm{n}}-
	\frac{1}{M^{2}}\tilde{\nabla}_{k}\hat{\varPi}_{k;\,\bm{n}},
\end{equation}
the change of variables from $\varPi_{B;\,\bm{n}}$ to $A^{(\mathrm{E})}_{4;\,\bm{n}}$ by
\begin{equation}
	A^{(\mathrm{E})}_{0;\,\bm{n}}=\varPi_{B;\,\bm{n}}+
	\frac{i}{M^{2}}\tilde{\nabla}_{k}\varPi_{k;\,\bm{n}}
\end{equation}
must be obtained as the relation of eigenvalues of related operators as
\begin{equation}
	\hat{A}_{0;\,\bm{n}}\vert\{\varPi\}\rangle=
	\left(i\varPi_{B;\,\bm{n}}-
	\frac{1}{M^{2}}\tilde{\nabla}_{k}\varPi_{k;\,\bm{n}}\right)
	\vert\{\varPi\}\rangle\equiv
	iA^{(\mathrm{E})}_{4;\,\bm{n}}\vert\{\varPi\}\rangle.
\end{equation}

Our final comment on the path integral considered here is that our prescription is available only under the trace formula because we have introduced spatial components of $A^{(\mathrm{E})}_{\mu;\,n}$ as a combination of eigenvalues of $\hat{U}_{k;\,\bm{n}}$ and $\hat{B}_{\bm{n}}$ in front of $\vert\{\varPhi\}\rangle$ while $\vert\{\varPi\}\rangle$ being in need to introduce $A^{(\mathrm{E})}_{4;\,n}$ as an integration variable of the path integral. Furthermore, to obtain an expression of the path integral in terms of $A^{(\mathrm{E})}_{\mu;\,n}$, the order of integrations are significant; without carrying out the integrations with respect to $\varPi_{k;\,n}$s and $B_{n}$s, we cannot arrive the desired path integral. Therefore the number of integrations with respect to $\varPhi_{I;\,n}$s and $\varPi_{I;\,n}$ must be valanced.

\section{Discussions}

Through the previous arguments, our definition of the path integral deeply depends on the Euclidean technique. Within the framework of Euclidean path integral, as far as the mass of the particle is kept to be $m^{2}>0$, the propagator is always well-defined. We may introduce $-i\epsilon$ to the denominator of the propagator just on putting $x_{4}=ix_{0}$ through the analytic continuation as a function of $x_{4}$ by keeping the Feynman's boundary condition. We here discuss the path integral formulated in the Minkowski space to examine whether we can properly regularize the path integral by Feynman's prescription, i.e. $m^{2}\to m^{2}-i\epsilon$. 

By putting $x_{4}=ix_{0}$ and therefore $\beta=iT=iL$, we can consider the Minkowski version of \eqref{eq:ZE}.
We again make the same change of variables given by \eqref{eq:change} as before, but we write $A_{0;\,n}$ instead of $A^{(\mathrm{E})}_{4;\,n}$, and carry out Fresnel integrals with respect to $\varPi_{k;\,n}-(\nabla_{0}A_{k;\,n}-i\nabla_{k}A_{0;\,n})$s and $B_{n}+(\tilde{\nabla}_{0}A_{0;\,n}+i\tilde{\nabla}_{k}A_{k;\,n}+iJ_{B;\,n})/\alpha$s, with keeping shifts in complex planes of $\varPi_{k;\,n}$s and $B_{n}$s finite, to obtain
\begin{equation}
	Z[J]=\frac{1}{\sqrt{\det\{(2\pi)^{4}\alpha\}}}
	\int_{-\infty}^{\infty}\!\!\prod_{n_{\mu}=-N}^{N}\prod_{\mu=0}^{3}
	dA_{\mu;\,n}e^{iS_{J}[A_{\mu}]},
\end{equation}
where the action of the path integral is given by
\begin{multline}
	S_{J}[A_{\mu}]=-\frac{1}{2}A_{0;\,n}\left\{
	\DAlambert+M^{2}-i\epsilon+\frac{1-\alpha}{\alpha}
	\nabla_{0}\tilde{\nabla}_{0}\right\}A_{0;\,n}-
	\frac{1}{2}A_{k;\,n}\left\{\delta_{k,\,l}\left(
	\DAlambert+M^{2}-i\epsilon\right)-\frac{1-\alpha}{\alpha}
	\nabla_{k}\tilde{\nabla}_{l}\right\}A_{l;\,n}\\
	-\frac{i(1-\alpha)}{2\alpha}\left(
	A_{0;\,n}\nabla_{0}\tilde{\nabla}_{k}A_{k;\,n}+
	A_{k;\,n}\nabla_{k}\tilde{\nabla}_{0}A_{0;\,n}\right)-
	i\tilde{J}_{0;\,n}A_{0;\,n}-\tilde{J}_{k;\,n}A_{k;\,n}-
	\frac{1}{2\alpha}J_{B;\,n}^{2}.
\end{multline}
Here $\DAlambert\equiv\nabla_{0}\tilde{\nabla}_{0}-\Laplace$ and modified sources are given by
\begin{equation}
	\tilde{J}_{0;\,n}\equiv
	J_{0;\,n}+\frac{1}{\alpha}\nabla_{0}J_{B;\,n},\quad
	\tilde{J}_{k;\,n}\equiv
	J_{k;\,n}-\frac{1}{\alpha}\nabla_{k}J_{B;\,n}.
\end{equation}
Note that we have introduced $-i\epsilon$ above to regularize the oscillatory integrals with the same sign for all components of the vector field. Thanks to the existence of $-i\epsilon$, we can carry out integrations with respect to $A_{\mu;\,n}$s as Gaussian integrals.

We introduce here $\tilde{A}_{\mu;\,n}$ by putting $\tilde{A}_{0;\,n}\equiv iA_{0;\,n}$ as well as $\tilde{A}_{k;\,n}\equiv A_{k;\,n}$ to rewrite the action above in a covariant form. To this aim we further introduce contravariant components to the sources and by writing $-\nabla_{k}=\nabla^{k}$, we rewrite the source terms as
\begin{equation}
	i\tilde{J}_{0;\,n}A_{0;\,n}+\tilde{J}_{k;\,n}A_{k;\,n}=
	\tilde{J}^{\mu;\,n}\tilde{A}_{\mu;\,n},\quad
	\tilde{J}^{\mu;\,n}\equiv
	J^{\mu}{}_{;\,n}+\frac{1}{\alpha}\nabla^{\mu}J_{B;\,n}.
\end{equation}
We then obtain
\begin{equation}
	S_{J}[A_{\mu}]=-\frac{1}{2}\tilde{A}_{\mu;\,n}
	\left(D^{-1}\right)^{\mu,\,\nu}
	\tilde{A}_{\nu;\,n}-
	\tilde{J}^{\mu;\,n}\tilde{A}_{\mu;\,n}-
	\frac{1}{2\alpha}J_{B;\,n}^{2},
\end{equation}
where
\begin{equation}
	\left(D^{-1}\right)^{\mu,\,\nu}\equiv
	-\eta^{\mu,\,\nu}\left(\DAlambert+M^{2}-i\epsilon\right)
	+\left(1-\frac{1}{\alpha}\right)\nabla^{\mu}\tilde{\nabla}^{\nu}.
\end{equation}
By completing the square of $\tilde{A}_{\mu;\,n}$ while keeping in mind the fact that Gaussian integrations must be carried out with respect to $A_{\mu;\,n}$, we obtain the generating functional
\begin{equation}
	Z[J]=Z_{0}e^{iW[J]},\quad
	Z_{0}\equiv\frac{1}{\sqrt{\det(\DAlambert+M^{2}-i\epsilon)^{3}
	\det(\DAlambert+\alpha M^{2}-i\epsilon)}},
\end{equation}
in which the generating functional $W[J]$ is given by
\begin{equation}
	W[J]=\frac{1}{2}\tilde{J}^{\mu}{}_{;\,n}
	D^{n,\,n{'}}_{\mu,\,\nu}\tilde{J}^{\nu}{}_{;\,n{'}}-
	\frac{1}{2\alpha}J_{B;\,n}^{2}.
\end{equation}
It is evident that the propagator for the vector field $A_{\mu;\,n}$ is given By
\begin{equation}
	D^{n,\,n{'}}_{\mu,\,\nu}=-\left(\eta_{\mu,\,\nu}+
	\frac{1}{M^{2}}\nabla_{\mu}\tilde{\nabla}_{\nu}\right)
	i\Delta_{\mathrm{F}}^{n,\,n{'}}+
	\frac{1}{M^{2}}\nabla_{\mu}\tilde{\nabla}_{\nu}
	i\Delta_{\mathrm{F}}^{(\alpha)\,n,\,n{'}},
\end{equation}
where $i\Delta_{\mathrm{F}}^{n,\,n{'}}$ and $i\Delta_{\mathrm{F}}^{(\alpha)\,n,\,n{'}}$ are defined by
\begin{equation}
	\left(\DAlambert+M^{2}-i\epsilon\right)
	i\Delta_{\mathrm{F}}^{n,\,n{'}}=\delta^{n,\,n{'}},\quad
	\left(\DAlambert+\alpha M^{2}-i\epsilon\right)
	i\Delta_{\mathrm{F}}^{(\alpha)\,n,\,n{'}}=\delta^{n,\,n{'}}.
\end{equation}

Classical solutions are determined by
\begin{equation}
	A^{\mathrm{cl.}}_{\mu;\,n}=D_{\mu,\,\nu}^{n,\,n{'}}
	\tilde{J}^{\nu}{}_{;\,n{'}},\quad
	B^{\mathrm{cl.}}_{n}=-\frac{1}{\alpha}\tilde{\nabla}^{\mu}
	D_{\mu,\,\nu}^{n,\,n{'}}
	\tilde{J}^{\nu}{}_{;\,n{'}}
	-\frac{1}{\alpha}J_{B;\,n}
\end{equation}
whose inversions yield
\begin{equation}
	\tilde{J}^{\mu}{}_{;\,n}=\left(D^{-1}\right)^{\mu,\,\nu}_{n,\,n{'}}
	A^{\mathrm{cl.}}_{\nu;\,n{'}},\quad
	J_{B;\,n}=-\alpha B_{n}-\tilde{\nabla}^{\mu}A^{\mathrm{cl.}}_{\mu;\,n{'}}.
\end{equation}
Taking these into account, we can easily rewrite the generating functional $W[J]$ in terms of classical solutions as
\begin{equation}
	W[J]=\frac{1}{2}A^{\mathrm{cl.}}_{\mu;\,n}
	\left(D^{-1}\right)^{\mu,\,\nu}_{n,\,n{'}}
	A^{\mathrm{cl.}}_{\nu;\,n{'}}-
	\frac{1}{2\alpha}\left(
	\alpha B_{n}+\tilde{\nabla}^{\mu}A^{\mathrm{cl.}}_{\mu;\,n}\right)^{2}.
\end{equation}
By remembering the explicit form of $\left(\tilde{D}^{-1}\right)^{\mu,\,\nu}_{n,\,n{'}}$, we obtain
\begin{equation}
	W[J]=
	\frac{1}{2}
	A^{\mathrm{cl.}}_{\mu;\,n}\left\{
	-\eta^{\mu,\,\nu}\left(\DAlambert+M^{2}-i\epsilon\right)+
	\nabla^{\mu}\tilde{\nabla}^{\nu}\right\}
	A^{\mathrm{cl.}}_{\nu;\,n}-
	B^{\mathrm{cl.}}_{n}\tilde{\nabla}^{\mu}A^{\mathrm{cl.}}_{\mu;\,n}-
	\frac{\alpha}{2}B^{\mathrm{cl.}\,2}_{n}.
\end{equation}
It will be straightforward to find that there holds 
$J^{\mu}{}_{;\,n}A^{\mathrm{cl.}}_{\mu;\,n}+J_{B;\,n}B^{\mathrm{cl.}}_{n}=2W[J]$. We thus, dropping $-i\epsilon$ now, obtain the effective action
\begin{equation}
	\varGamma[A^{\mathrm{cl.}}_{\mu},\,B^{\mathrm{cl.}}]=
	-\frac{1}{4}F^{\mu\,\nu}{}_{;\,n}F_{\mu\,\nu;\,n}+
	\frac{1}{2}M^{2}A^{\mathrm{cl.}}_{\mu;\,n}A^{\mathrm{cl.}\,\mu}_{;\,n}+
	B^{\mathrm{cl.}}_{n}\tilde{\nabla}^{\mu}A^{\mathrm{cl.}}_{\mu;\,n}+
	\frac{\alpha}{2}B^{\mathrm{cl.}\,2}_{n}
\end{equation}
which is nothing but the classical action for the system we have defined at the beginning of this paper.
Thus we have been convinced that we can regularize the oscillatory path integral in the Minkowski space by the introduction of Feynman's $-i\epsilon$ in a proper manner. The fact that the $-i\epsilon$ prescription works fine with our formulation of the Minkowski path integral on the basis of the indefinite metric Hilbert space to yield the correct propagator is equivalent to that the path integral so defined fulfills the Euclidicity postulate\cite{Abers-Lee,Coleman:85}. In this regard, since the path integral considered here is obtained just by putting $x_{4}=ix_{0}$ in the Euclidean version considered in the previous section in terms of $A^{(\mathrm{E})}_{\mu;\,n}$, we can translate results obtained there into the ones for Minkowski space; though we have written the same object as $A_{0;\,n}$ which was written as $A_{4;\,n}$ in the Euclidean case. Hence some non-perturbative results, such as the instanton effect, in the Euclidean formulation may have their Minkowski counterparts as long as the path integral is regularized by the correct $-i\epsilon$ prescription though the action of the path integral may lose the covariance.

It will be interesting to compare the process above with the covariant formulation of the path integral for the system under consideration.
On the basis of canonical commutation relations
\begin{equation}
\label{eq:CCR}
	[\hat{A}_{\mu;\,\bm{n}},\,\hat{\varPi}^{\nu}_{;\,\bm{n}{'}}]=
	i\delta_{\mu}^{\nu}\delta_{\bm{n},\,\bm{n}{'}},
\end{equation}
where $\hat{\varPi}^{k}_{;\,\bm{n}}=-\hat{F}^{0\,k}{}_{;\,\bm{n}}$ and $\hat{\varPi}^{0}_{\bm{n}}=\hat{B}_{\bm{n}}$, it is usually expected to be able to formulate path integrals on a Hilbert space with the positive definite metric. In this case, we assume the existence of eigenvectors $\vert\{A_{\mu}\}\rangle$ as well as $\vert\{\varPi^{\mu}\}\rangle$ to fulfill
\begin{equation}
	\hat{A}_{\mu;\,\bm{n}}\vert\{A_{\mu}\}\rangle=
	\vert\{A_{\mu}\}\rangle
	A_{\mu;\,\bm{n}},\quad
	\hat{\varPi}^{\mu}{}_{;\,\bm{n}}\vert\{\varPi^{\mu}\}\rangle=
	\vert\{\varPi^{\mu}\}\rangle
	\varPi^{\mu}{}_{;\,\bm{n}}
\end{equation}
and the completeness of these eigenvectors.
We will be then able to formulate a path integral for our system to describe the generating functional to be given by
\begin{multline}
\label{eq:gf05}
	Z{'}[J]=
	\frac{1}{(2\pi)^{4(2N+1)^{4}}}
	\int_{-\infty}^{\infty}\!\!\prod_{n_{\mu}=-N}^{N}\prod_{\mu=0}^{3}
	d\varPi^{\mu}{}_{;\,n}\,dA_{\mu;\,n}
	\exp\left[\vphantom{\frac{M^{2}}{2}}
	i\varPi^{k}{}_{;\,n}\nabla_{0}A_{k;\,n}+iB_{n}\tilde{\nabla}^{0}A_{0;\,n}
	\right.\\
	\left.
	-i\left\{
	\frac{1}{2}\varPi^{k\,2}_{;\,n}
	+\varPi^{k}{}_{;\,n}\nabla_{k}A_{0;\,n}
	+\frac{1}{4}F^{k\,l}{}_{;\,n}F_{k\,l;\,n}-
	\frac{M^{2}}{2}A_{\mu;\,n}A^{\mu}{}_{;\,n}-
	B_{n}\tilde{\nabla}^{k}A_{k;\,n}-
	\frac{\alpha}{2}B^{2}_{n}+
	J^{\mu}{}_{;\,n}A_{\mu;\,n}+J_{B;\,n}B_{n}\right\}\right],
\end{multline}
where we have written $\varPi^{0}{}_{;\,n}$ as $B_{n}$ in the exponent. 
We have assumed the use of the completeness of the eigenvector of $\hat{A}_{0;\,\bm{n}}$ in combination with that of $\hat{\varPi}^{k}{}_{;\,\bm{n}}$ so that we are able to obtain the term $iB_{n}\tilde{\nabla}^{0}A_{0;\,n}$ in the first line. If we have made the use of the completeness together with that of $\hat{A}_{k;\,\bm{n}}$, this term must be $iB_{n}\nabla^{0}A_{0;\,n}$, instead. The difference between these two situations will not be the matter in the continuum limit, it is, however, important to make the action be covariant on the lattice. 

By carrying out integrations with respect to $\varPi^{k}{}_{;\,n}$ and $B_{n}$, we obtain
\begin{equation}
	Z{'}[J]=\frac{1}{\{(2\pi)^{2}i\sqrt{\alpha}\}^{(2N+1)^{4}}}
	\int_{-\infty}^{\infty}\!\!\prod_{n_{\mu}=-N}^{N}\prod_{\mu=0}^{3}
	dA_{\mu;\,n}
	e^{iS{'}_{J}[A_{\mu}]},
\end{equation}
where the action of the path integral is given by
\begin{equation}
	S{'}_{J}[A_{\mu}]\equiv
	-\frac{1}{2}A_{\mu;\,n}
	\left\{-\eta^{\mu,\,\nu}\left(\DAlambert+M^{2}\right)-
	\frac{1-\alpha}{\alpha}\nabla^{\mu}\tilde{\nabla}^{\nu}\right\}
	A_{\nu;\,n}-\tilde{J}^{\mu}{}_{;\,n}A_{\mu;\,n}-
	\frac{1}{2\alpha}J^{2}_{B;\,n},
\end{equation}
in which modified sources being the same as before.

If we carry out the Fresnel integrals with respect to $A_{\mu;\,n}$s, we may obtain, as a result of the covariant path integral above, the generating functional $Z{'}[J]$ to be given by
\begin{equation}
	Z{'}[J]=Z{'}_{0}e^{iW{'}[J]},\quad
	Z{'}_{0}\equiv\frac{1}
	{\sqrt{\det\left(\DAlambert+M^{2}\right)^{3}
	\det\left(\DAlambert+\alpha M^{2}\right)}},
\end{equation}
where the generating functional $W{'}[J]$ is given by
\begin{equation}
	W{'}[J]\equiv
	\frac{1}{2}\tilde{J}^{\mu}{'}_{;\,n}D_{\mu,\,\nu}^{n,\,n{'}}
	\tilde{J}^{\nu}{'}_{;\,n{'}}-
	\frac{1}{2\alpha}J^{2}_{B;\,n},
\end{equation}
with the definition
\begin{equation}
	D_{\mu,\,\nu}^{n,\,n{'}}\equiv
	-\left(\eta_{\mu,\,\nu}+
	\frac{1}{M^{2}}\nabla_{\mu}\tilde{\nabla}_{\nu}\right)
	\left(\DAlambert+M^{2}\right)^{-1}+
	\frac{1}{M^{2}}\nabla_{\mu}\tilde{\nabla}_{\nu}
	\left(\DAlambert+\alpha M^{2}\right)^{-1}.
\end{equation}
By defining classical solutions to satisfy
\begin{equation}
	A^{\mathrm{cl.}}{'}_{\mu;\,n}\equiv
	\frac{\partial W{'}[J]}{\partial J^{\mu}{}_{n}},\quad
	B^{\mathrm{cl.}}{'}_{n}\equiv
	\frac{\partial W{'}[J]}{\partial J_{B;\,n}},
\end{equation}
we may finally obtain the effective action
\begin{equation}
	\varGamma{'}[A^{\mathrm{cl.}}{'}_{\mu},B^{\mathrm{cl.}}{'}_{n}]=
	-\frac{1}{2}
	A^{\mathrm{cl.}}{'}_{\mu;\,n}\left\{
	\eta^{\mu,\nu}\left(\DAlambert+M^{2}\right)
	-\nabla^{\mu}\tilde{\nabla}^{\nu}\right\}
	A^{\mathrm{cl.}}{'}_{\nu;\,n}
	-B^{\mathrm{cl.}\,'}_{n}\tilde{\nabla}^{\mu}A^{\mathrm{cl.}}{'}_{\mu;\,n}
	-\frac{\alpha}{2}B^{\mathrm{cl.}\,'\,2}_{n}
\end{equation}
as the result of covariant path integral in the Minkowski space.

The manipulation above must be, however, corrected to be acceptable in following points. First of all, Fresnel integrals with respect to $A_{\mu;\,n}$s must be treated more carefully; if we Fourier transform $A_{\mu;\,n}$ by expanding into the series of $F_{n_{\mu}}^{r_{\mu}}$, the spectrum of $\DAlambert+M^{2}$ becomes $-P_{\mu;\,r_{\mu}}P^{\mu\,*}_{\ \ ;\,r_{\mu}}+M^{2}$ hence we need to divide the treatment of Fresnel integral depending the sign of $-P_{\mu;\,r_{\mu}}P^{\mu\,*}_{\ \ ;\,r_{\mu}}+M^{2}$.
Secondly, we cannot defne the inverse of $\DAlambert+M^{2}$ even on the lattice, if it happens to occur $-P_{\mu;\,r_{\mu}}P^{\mu\,*}_{\ \ ;\,r_{\mu}}+M^{2}=0$ and, of course, for that case the Fresnel integral becomes ill-defined.
To avoid such troubles, we need to regularize the Fresnel integrals by introducing $-i\epsilon$ in a proper manner. Clearly, we need the following replacement
\begin{equation}
	A_{\mu;\,n}
	\eta^{\mu,\,\nu}\left(\DAlambert+M^{2}\right)A_{\nu;\,n}\to
	A_{0;\,n}
	\left(\DAlambert+M^{2}+i\epsilon\right)A_{0;\,n}-
	A_{k;\,n}\delta^{k,\,l}
	\left(\DAlambert+M^{2}-i\epsilon\right)A_{l;\,n}
\end{equation}
to make the Fresnel integrals to be Gaussian ones. This will, however, immediately results in the breaking of the covariance as well as the causality. The origin of this discrepancy will be attributed to the use of the positive metric Hilbert space for the representation of the canonical commutation relations \eqref{eq:CCR}. Namely, the definition of the creation and annihilation operators for the mode quantized with indefinite metric such as $\hat{A}_{0;\,\bm{n}}$ in Feynman gauge ($\alpha=1$), for example, will  necessarily be interchanged between the positive norm representation and the indefinite metric case. We have therefore observed that the manifestly covariant path integral formulated on the basis of positive norm representation fails to yield really covariant result; it rather results in non-covariant effective action and also breaks the causality. On this point, to our knowledge, we cannot find any reports in the literature excepting the one\cite{Arisue:81} by Arisue {\em et al.\/} Boulware and Gross\cite{Boulware-Gross} discuss about the need of indefinite metric representation in connection with the Lee-Wick ghost vector field but not for the usual vector field. In Ref.~\cite{Kashiwa:81} and Ref.~\cite{Kashiwa-Sakamoto}, the authors make assertions that they can formulate path integrals for vector fields without the use of negative norm Hilbert space; but, as is already commented earlier in this paper, the Gaussian identities they have utilized involves the equivalent which explains why they were able to formulate Euclidean path integrals of vector fields. In other words, the use of the indefinite metric Hilbert space is inevitable in the Euclidean path integral for the vector field. In many textbook  on the quantum field theory, however, this point seems to be simply ignored and in the denominator of the Feynman propagator $-i\epsilon$ appears regardless of the component.
This is identical to carry out Fresnel integrals above without regularization and insert $-i\epsilon$ afterwise. The covariant path integral defined in such a way cannot, of course, be converted into the imaginary time by simply putting $ix_{0}=x_{4}$. Thus results in the introduction of the Euclidean vector field by writing  $A_{0;\,n}=iA_{4;\,n}$ as a prescription to obtain a convergent path integral.
It is the use of indefinite metric Hilbert space that provides the correct way of defining path integrals of vector fields. In this regard, the $-i\epsilon$ prescription in the Minkowski case is almost equivalent to the Euclidean technique to fulfill the Euclidicity postulate as long as the use of the indefinite metric Hilbert space is concerned.

\section{Conclusion}
We have introduced a unified method of finding eigenvectors of field operators in the first half of this paper to obtain eigenvectors of a massive vector field accompanied with a scalar field as well as the ones for their canonical conjugates basing on the Lagrangian proposed by Nakanishi\cite{Nakanishi:72}.
Since our method is on the basis of covariant canonical formalism, there naturally appears the degree of freedom that requires indefinite metric Hilbert space for quantization, i.e. $B$-field. Eigenvalues of $B$-field and its canonical conjugate become pure imaginary. Although a similar result, though restricted to the Feynman gauge, was already given by Arisue {\em et al.\/} in Ref.~\cite{Arisue:81}, our method given in this paper may have advantage in the sense that we can systematically perform the calculations needed in finding eigenvectors of field operators. Moreover, in the course of finding eigenvectors, our method naturally provides resolutions of unity expressed in terms of these eigenvectors.

Resolutions of unity thus obtained can be utilized to formulate path integrals for the vector field keeping good connection with the covariant canonical formalism.
On the basis of these fundamental ingredients, we have considered path integrals of the vector field in Euclidean space and in Minkowski space as well in the second half of this paper. By the fact that operators quantized with indefinite metric require imaginary eigenvalues, the Euclidean path integrals of the system considered in this paper become well-defined but the manifest covariance of the action in the exponent of the path integral cannot be expected by the same reason in the Minkowski case. We have needed to introduce $\tilde{A}_{\mu;\,n}$, in which $\tilde{A}_{0;\,n}=iA_{0;\,n}$ being a fake variable, in order to write the action in a covariant manner. On the other hand, the Euclidean path integral was naturally transformed into the manifestly covariant form expressed in terms of the Euclidean vector field $A^{(\mathrm{E})}_{\mu;\,n}$. The emergnce of $iA^{(\mathrm{E})}_{4;\,n}$ as the eigenvalue of $\hat{A}_{0;\,\bm{n}}$ is the consequence of the use of indefinite metric representation for the canonical commutation relations. In this connection we must emphasize that the path integral in terms of the Euclidean vector field cannot be interpreted as the one obtained by the Wick rotation $A_{0;\,n}=iA_{4;\,n}$ from the manifestly covariant one in Minkowski space; it should be rather reinterpreted as the consequence of the indefinite metric representation in the covariant canonical formalism. On the basis of this explanation, we can understand the meaning of the Euclidean path integral of the vector field and results there obtained can be naturally translated into the Minkowski ones.
The manifestly covariant path integral that occupies the literature cannot be regularized in a covariant manner by the use of Feynman's $-i\epsilon$ prescription on the contrary; i.e. it does not support the Euclidicity postulate. The effective action defined by the path integral with non-covariant action in Minkowski space possesses the manifest covariance; the absence of the manifest covariance in the action of a path integral does not necessarily mean, once the use of indefinite metric Hilbert space is accepted, the loss of covariance in its results and therefore the covariance of a path integral must be related to its results such as the effective action or Green's functions. Since we have obtained the covariant propagator from our formulation of the path integral, the use of indefinite metric representation as its background seems to be the key for a path integral of the vector field to fulfill the Euclidean postulate.

We may generalize the method of constructing field diagonal representation shown in this paper to other case such as the quantization of Lee-Wick models with indefinite metric representation. Euclidicity postulate may work as the guiding principle to formulate Euclidean path integrals even for such systems. This will be discussed elsewhere.

We have no chance to discuss the BRS invariance of the system because our system explicitly breaks the BRS symmetry due to the mass term of the vector field. It is, however, possible to extend the system to be a part of the Higgs model\cite{Higgs} with a covariant gauge fixing. We will be then able to discuss the BRS invariance of the system. Due to the imaginary eigenvalues of the degree with indefinite metric Hilbert space, the action of the path integral in Minkowski space loses manifest covariance; we may obtain a non-covariant action also for the Faddeev-Popov ghost. On the other hand, since the effective action of that path integral is covariant, the corresponding part of the ghost may possess covariance. In this regard, we may further extend our system to include the non-Abelian gauge field and utilize our results in this paper to describe the free field of such systems and formulate the perturbative expansion. These are future tasks for us.

\appendix
\section{Eigenvectors of position and momentum operators}
\subsection{Positive norm case}
Let us begin with a Lagrangian
\begin{equation}
	L=\frac{\ 1\ }{\ 2\ }\dot{q}^{2}-\frac{\ 1\ }{\ 2\ }q^{2}.
\end{equation}
The corresponding Hamiltonian
\begin{equation}
	\hat{H}=\frac{\ 1\ }{\ 2\ }\hat{p}^{2}+\frac{\ 1\ }{\ 2\ }\hat{q}^{2}
\end{equation}
can be rewritten as
\begin{equation}
	\hat{H}=\hat{a}^{\dagger}\hat{a}+\frac{\ 1\ }{\ 2\ }
\end{equation}
if we introduce creation and annihilation operators by
\begin{equation}
	\hat{a}=\frac{\ 1\ }{\ \sqrt{2\,}\ }(\hat{q}+i\hat{p}),\quad
	\hat{a}^{\dagger}=\frac{\ 1\ }{\ \sqrt{2\,}\ }(\hat{q}-i\hat{p})
\end{equation}
to fulfill the commutation relation
\begin{equation}
	[\hat{a},\,\hat{a}^{\dagger}]=\bm{1}
\end{equation}
where $\bm{1}$ designates unity.
The vacuum $\vert0\rangle$ is defined by
\begin{equation}
	\hat{a}\vert0\rangle=0,\quad
	\langle0\vert0\rangle=1.
\end{equation}

To find the eigenvector of the position operator $\hat{q}$, we consider
\begin{equation}
\label{eq:q-projector01}
	\frac{\ 1\ }{\ 2\pi\ }\int_{-\infty}^{\infty}\!\!d\lambda
	e^{i\lambda(\hat{q}-q)}.
\end{equation}
By expressing $\hat{q}$ in terms of $\hat{a}$ and $\hat{a}^{\dagger}$, we rewrite $e^{i\lambda\hat{q}}$ as
\begin{equation}
	e^{i\lambda\hat{q}}=e^{-\lambda^{2}/4}
	e^{i\lambda\hat{a}^{\dagger}/\sqrt{2\,}}
	e^{i\lambda\hat{a}/\sqrt{2\,}}
\end{equation}
and sandwich this expression between two coherent states
\begin{equation}
	\langle z\vert=\langle 0\vert e^{z^{*}\hat{a}}\quad\text{and}\quad
	\vert z{'}\rangle=e^{\hat{a}^{\dagger}z{'}}\vert 0\rangle
\end{equation}
to find
\begin{equation}
	\langle z\vert e^{i\lambda\hat{q}}\vert z{'}\rangle=
	\langle z\vert z{'}\rangle
	e^{-\lambda^{2}/4+i\lambda(z^{*}+z{'})/\sqrt{2\,}}.
\end{equation}
Then, by carrying out the integration with respect to $\lambda$, we obtain
\begin{equation}
\label{eq:q-projector02}
	\frac{\ 1\ }{\ 2\pi\ }\int_{-\infty}^{\infty}\!\!d\lambda
	\langle z\vert e^{i\lambda(\hat{q}-q)}\vert z{'}\rangle=
	\langle z\vert z{'}\rangle
	\frac{\ 1\ }{\ \sqrt{\pi\,}\ }e^{-\{q-(z^{*}+z{'})/\sqrt{2\,}\}^{2}}.
\end{equation}
If we remember the following relations:
\begin{equation}
	\hat{a}\vert z\rangle=\vert z\rangle z,\ 
	\langle z\vert\hat{a}^{\dagger}=z^{*}\langle z\vert,\ 
	\langle z\vert0\rangle=\langle 0\vert z\rangle=1,\ 
	\langle z\vert z{'}\rangle=e^{z^{*}z{'}},
\end{equation}
we can rewrite the result above as
\begin{equation}
\label{eq:q-projector03}
\begin{aligned}
	&\frac{\ 1\ }{\ 2\pi\ }\int_{-\infty}^{\infty}\!\!d\lambda
	\langle z\vert e^{i\lambda(\hat{q}-q)}\vert z{'}\rangle\\
	=&\frac{\ 1\ }{\ \sqrt{\pi\,}\ }
	\langle z\vert
	\exp\left(-\frac{\ 1\ }{\ 2\ }(\hat{a}^{\dagger})^{2}+
	\sqrt{2\,}q\hat{a}^{\dagger}-\frac{\ 1\ }{\ 2\ }q^{2}\right)\vert0\rangle\\
	&\times
	\langle0\vert
	\exp\left(-\frac{\ 1\ }{\ 2\ }\hat{a}^{2}+
	\sqrt{2\,}q\hat{a}-\frac{\ 1\ }{\ 2\ }q^{2}\right)
	\vert z{'}\rangle.
\end{aligned}
\end{equation}
Integration of the right hand side of \eqref{eq:q-projector02} with respect to $q$ clearly shows that there holds
\begin{equation}
\label{eq:q-projector04}
	\frac{\ 1\ }{\ 2\pi\ }\int_{-\infty}^{\infty}\!\!dq\,d\lambda
	e^{i\lambda(\hat{q}-q)}=\bm{1}.
\end{equation}
Therefore we can write
\begin{equation}
	\frac{\ 1\ }{\ 2\pi\ }\int_{-\infty}^{\infty}\!\!d\lambda
	e^{i\lambda(\hat{q}-q)}=\vert q\rangle\langle q\vert
\end{equation}
in which $\vert q\rangle$ and $\langle q\vert$ are given by
\begin{equation}
	\vert q\rangle=\frac{\ 1\ }{\ \pi^{1/4}\ }
	\exp\left(-\frac{\ 1\ }{\ 2\ }(\hat{a}^{\dagger})^{2}+
	\sqrt{2\,}q\hat{a}^{\dagger}-\frac{\ 1\ }{\ 2\ }q^{2}\right)\vert0\rangle
\end{equation}
and
\begin{equation}
	\langle q\vert=\frac{\ 1\ }{\ \pi^{1/4}\ }\langle0\vert
	\exp\left(-\frac{\ 1\ }{\ 2\ }\hat{a}^{2}+
	\sqrt{2\,}q\hat{a}-\frac{\ 1\ }{\ 2\ }q^{2}\right).
\end{equation}
It is straightforward to check that these are eigenvectors of $\hat{q}$ to satisfy
\begin{equation}
	\hat{q}\vert q\rangle=\vert q\rangle q\quad\text{and}\quad
	\langle q\vert\hat{q}=q\langle q\vert.
\end{equation}

In the same way, we can find that eigenvectors of the momentum operator $\hat{p}$ are given by
\begin{equation}
	\vert p\rangle=\frac{\ 1\ }{\ \pi^{1/4}\ }
	\exp\left(\frac{\ 1\ }{\ 2\ }(\hat{a}^{\dagger})^{2}+
	i\sqrt{2\,}p\hat{a}^{\dagger}-\frac{\ 1\ }{\ 2\ }p^{2}\right)\vert0\rangle
\end{equation}
and
\begin{equation}
	\langle p\vert=\frac{\ 1\ }{\ \pi^{1/4}\ }\langle0\vert
	\exp\left(\frac{\ 1\ }{\ 2\ }\hat{a}^{2}-
	i\sqrt{2\,}p\hat{a}-\frac{\ 1\ }{\ 2\ }p^{2}\right).
\end{equation}
Inner products of the eigenvectors obtained above are given by
\begin{equation}
	\langle q\vert q{'}\rangle=\delta(q-q{'}),\
	\langle p\vert p{'}\rangle=\delta(p-p{'})\ \text{and}\ 
	\langle q\vert p\rangle=\frac{\ 1\ }{\ \sqrt{2\pi\,}\ }e^{ipq}.
\end{equation}
The completeness of these eigenvectors are written as
\begin{equation}
	\int_{-\infty}^{\infty}\!\!\vert q\rangle\langle q\vert\,dq=\bm{1}
	\ \text{and}\ 
	\int_{-\infty}^{\infty}\!\!\vert p\rangle\langle p\vert\,dp=\bm{1}.
\end{equation}
These are the fundamental ingredients for constructing path integrals by means of time slicing method.

\subsection{Negative norm case}
Quantization of a system described by a Lagrangian
\begin{equation}
	L=-\frac{\ 1\ }{\ 2\ }\dot{q}^{2}+\frac{\ 1\ }{\ 2\ }q^{2}
\end{equation}
can be achieved by constructing the representation of the canonical commutation relation on the Hilbert space with indefinite metric.
The Hamiltonian
\begin{equation}
	\hat{H}=-\frac{\ 1\ }{\ 2\ }\hat{p}^{2}-\frac{\ 1\ }{\ 2\ }\hat{q}^{2}
\end{equation}
is rewritten as
\begin{equation}
	\hat{H}=-\hat{a}^{\dagger}\hat{a}+\frac{\ 1\ }{\ 2\ }
\end{equation}
if we introduce creation and annihilation operators by
\begin{equation}
	\hat{a}=\frac{\ 1\ }{\ \sqrt{2\,}\ }(\hat{q}-i\hat{p}),\quad
	\hat{a}^{\dagger}=\frac{\ 1\ }{\ \sqrt{2\,}\ }(\hat{q}+i\hat{p})
\end{equation}
to fulfill the commutation relation
\begin{equation}
	[\hat{a},\,\hat{a}^{\dagger}]=-\bm{1}.
\end{equation}
Note that the Heisenberg operator $\hat{a}(t)=e^{i\hat{H}t}\hat{a}e^{-i\hat{H}t}$ behaves as $\hat{a}(t)=e^{-it}\hat{a}$. This is the reason why we define the annihilation operator this way.
To construct the representation of the algebra, we define the vacuum $\vert0\rangle$ to satisfy $\hat{a}\vert0\rangle=0$.
The eigenvector of the number operator $\hat{N}=\hat{a}^{\dagger}\hat{a}$ is then given by
\begin{equation}
	\vert n\rangle=\frac{\ (\hat{a}^{\dagger})^{n}\ }{\ \sqrt{n!\,}\ }
	\vert 0\rangle
\end{equation}
and fulfills
\begin{equation}
	\hat{N}\vert n\rangle=-n\vert n\rangle.
\end{equation}

Due to the commutation relation above, the norm of these eigenvectors are not always positive and given by
\begin{equation}
	\langle n\vert n{'}\rangle=(-1)^{n}\delta_{n\,n{'}}.
\end{equation}
To remove $(-1)^{n}$ above, we introduce the metric operator by $\hat{\eta}=e^{i\pi\hat{N}}$ and define
\begin{equation}
	\langle\underline{n}\vert=\langle n\vert\hat{\eta}
\end{equation}
as the conjugate of $\vert n\rangle$.
Then the sum
$\sum_{n=0}^{\infty}\vert n\rangle\langle\underline{n}\vert$,
when multiplied to $\vert n{'}\rangle$, reproduces $\vert n{'}\rangle$.
We can therefore regard it as the resolution of unity:
\begin{equation}
	\sum_{n=0}^{\infty}\vert n\rangle\langle\underline{n}\vert=\bm{1}.
\end{equation}
The coherent state defined by $\vert z\rangle=e^{-\hat{a}^{\dagger}z}\vert0\rangle$ together with its conjugate $\langle\underline{z}\vert=\langle0\vert e^{z^{*}\hat{a}}$ forms a complete set and yields the resolution of unity
\begin{equation}
\label{eq:n-resolution01}
	\int\!\!\frac{\ dz\,dz^{*}\ }{\pi}e^{-z^{*}z}
	\vert z\rangle\langle\underline{z}\vert=\bm{1},
\end{equation}
where the integrations are done with respect to $\Re(z)$ and $\Im(z)$.
Note that they obey
\begin{equation}
	\hat{a}\vert z\rangle=\vert z\rangle z\quad\text{and}\quad
	\langle\underline{z}\vert\hat{a}^{\dagger}=-z^{*}\langle\underline{z}\vert.
\end{equation}

Similar to the positive norm case, we here consider
\begin{equation}
	\label{eq:nq-projector01}
	\frac{\ 1\ }{\ 2\pi\ }\int_{-\infty}^{\infty}\!\!d\lambda
	e^{\lambda(\hat{q}-iq)}
\end{equation}
and rewrite $e^{\lambda\hat{q}}$ as
\begin{equation}
	e^{\lambda\hat{q}}=e^{-\lambda^{2}/4}
	e^{\lambda\hat{a}^{\dagger}/\sqrt{2\,}}
	e^{\lambda\hat{a}/\sqrt{2\,}}
\end{equation}
to obtain
\begin{equation}
\label{eq:nq-projector02}
	\frac{\ 1\ }{\ 2\pi\ }\int_{-\infty}^{\infty}\!\!d\lambda
	\langle\underline{z}\vert e^{\lambda(\hat{q}-iq)}\vert z{'}\rangle=
	\langle\underline{z}\vert z{'}\rangle
	\frac{\ 1\ }{\ \sqrt{\pi\,}\ }e^{-\{q-i(z^{*}-z{'})/\sqrt{2\,}\}^{2}}.
\end{equation}
Integration of the right hand side above clearly exhibits that there holds
\begin{equation}
	\frac{\ 1\ }{\ 2\pi\ }\int_{-\infty}^{\infty}\!\!dq\,d\lambda
	e^{\lambda(\hat{q}-iq)}=\bm{1}.
\end{equation}
On this observation, we find that the eigenvector of $\hat{q}$ is given by
\begin{equation}
	\vert q\rangle=\frac{\ 1\ }{\ \pi^{1/4}\ }
	\exp\left(\frac{\ 1\ }{\ 2\ }(\hat{a}^{\dagger})^{2}-
	i\sqrt{2\,}q\hat{a}^{\dagger}-\frac{\ 1\ }{\ 2\ }q^{2}\right)\vert0\rangle
\end{equation}
together with its conjugate
\begin{equation}
	\langle\underline{q}\vert=\langle-q\vert=\frac{\ 1\ }{\ \pi^{1/4}\ }\langle0\vert
	\exp\left(\frac{\ 1\ }{\ 2\ }\hat{a}^{2}-
	i\sqrt{2\,}q\hat{a}-\frac{\ 1\ }{\ 2\ }q^{2}\right).
\end{equation}
The eigenvalue of $\hat{q}$ is pure imaginary because it satisfies
\begin{equation}
	\hat{q}\vert q\rangle=iq\vert q\rangle\quad\text{and}\quad
	\langle\underline{q}\vert\hat{q}=iq\langle\underline{q}\vert.
\end{equation}

In the same way, we find that the eigenvector of $\hat{p}$ is given by
\begin{equation}
	\vert p\rangle=\frac{\ 1\ }{\ \pi^{1/4}\ }
	\exp\left(-\frac{\ 1\ }{\ 2\ }(\hat{a}^{\dagger})^{2}+
	\sqrt{2\,}p\hat{a}^{\dagger}-\frac{\ 1\ }{\ 2\ }p^{2}\right)\vert0\rangle
\end{equation}
together with its conjugate
\begin{equation}
	\langle\underline{p}\vert=\langle-p\vert=\frac{\ 1\ }{\ \pi^{1/4}\ }\langle0\vert
	\exp\left(-\frac{\ 1\ }{\ 2\ }\hat{a}^{2}-
	\sqrt{2\,}p\hat{a}-\frac{\ 1\ }{\ 2\ }p^{2}\right).
\end{equation}
The eigenvalue of $\hat{p}$ is also pure imaginary and given by
\begin{equation}
	\hat{p}\vert p\rangle=-ip\vert p\rangle\quad\text{and by}\quad
	\langle\underline{p}\vert\hat{p}=-ip\langle\underline{p}\vert.
\end{equation}
Inner products of the eigenvectors thus obtained are given by
\begin{equation}
	\langle\underline{q}\vert q{'}\rangle=\delta(q-q{'}),\
	\langle\underline{p}\vert p{'}\rangle=\delta(p-p{'})\ \text{and}\ 
	\langle\underline{q}\vert p\rangle=\frac{\ 1\ }{\ \sqrt{2\pi\,}\ }e^{ipq}.
\end{equation}
The completeness of these eigenvectors are written as
\begin{equation}
	\int_{-\infty}^{\infty}\!\!\vert q\rangle\langle\underline{q}\vert\,dq=
	\bm{1}
	\ \text{and}\ 
	\int_{-\infty}^{\infty}\!\!\vert p\rangle\langle\underline{p}\vert\,dp=
	\bm{1}.
\end{equation}

\subsection{Euclidean path integrals}
We make use of expressions above to formulate a time sliced path integral for the Hamiltonian given at the top of this subsection. To do so, we first consider the short time kernel
\begin{equation}
\label{eq:n-kernel01}
	\langle\underline{q}\vert\left(\bm{1}-\epsilon\hat{H}\right)
	\vert q{'}\rangle=
	\int_{-\infty}^{\infty}\!\!dp\,
	\exp\left\{ip(q-q{'})-\epsilon\left(
	\frac{\ 1\ }{\ 2\ }p^{2}+\frac{\ 1\ }{\ 2\ }\bar{q}^{2}
	\right)\right\}
\end{equation}
for an infinitesimally small imaginary time interval $\epsilon$. Here, we have employed the mid-point prescription and wrote $\bar{q}\equiv(q+q^{'})/2$.
Note that $\hat{p}^{2}$ and $\hat{q}^{2}$ become negative since there eigenvalues are pure imaginary. Integration with respect to $p$ yields 
\begin{equation}
\label{eq:n-kernel02}
	\langle\underline{q}\vert\left(\bm{1}-\epsilon\hat{H}\right)
	\vert q{'}\rangle=
	\frac{1}{\ \sqrt{2\pi\epsilon\,}\ }
	\exp\left\{-\frac{1}{\ 2\epsilon\ }(q-q{'})^{2}-
	\frac{\ \epsilon\ }{\ 2\ }\bar{q}^{2}\right\}.
\end{equation}
We now set $\epsilon=\beta/N$ for a large integer $N$ and connecting $N$ pieces of the short time kernel above we obtain a time sliced path integral
\begin{equation}
\label{eq:n-kernel03}
\begin{aligned}
	&K(q_{\mathrm{F}},q_{\mathrm{I}};\beta)\\
	=&
	\lim\limits_{N\to\infty}\frac{1}{\ (2\pi\epsilon)^{N/2}\ }
	\int_{-\infty}^{\infty}\!\!\prod_{i=1}^{N-1}dq_{i}\,
	\exp\left[-\frac{\ \epsilon\ }{\ 2\ }\sum_{j=1}^{N}\left\{
	\left(\frac{\ \Delta q_{j}\ }{\epsilon}\right)^{2}+
	\frac{\ 1\ }{\ 2\ }\bar{q}_{j}^{2}\right\}\right]
\end{aligned}
\end{equation}
for the Feynman kernel with an imaginary time $\beta$. Here, we have set $q_{N}=q_{\mathrm{F}}$ and $q_{0}=q_{\mathrm{I}}$.
The form of this time sliced path integral is completely equivalent to the one for a usual harmonic oscillator. Therefore, when compared with the classical action, we find the opposite sign in the action in the imaginary time for a path integral of the negative oscillator. It is of course the consequence of imaginary eigenvalues of $\hat{p}$ and $\hat{q}$. The path integral above can be fully carried out to result in
\begin{equation}
\label{eq:n-kernel04}
	K(q,q{'};\beta)=
	\frac{1}{\ \sqrt{2\pi\sinh\beta\,}\ }
	\exp\left[-\frac{1}{\ 2\sinh\beta\ }\left\{
	(q^{2}+q{'}{}^{2})\cosh\beta-2qq{'}\right\}\right]
\end{equation}
as is wellknown for the positive harmonic oscillator.

Let us now consider the effect of an external source. To this aim, we introduce $\hat{a}^{\dagger}\eta(t)+\eta^{*}(t)\hat{a}$ for the imaginary time $t$ in $0<t<\beta$ to be added to the Hamiltonian $H=-\hat{a}^{\dagger}\hat{a}$.
We then define a generating functional given by
\begin{equation}
	Z_{-}[\eta^{*},\,\eta]=\langle\underline{z_{\mathrm{F}}}\vert
	\mathrm{T}\exp\left[-\int_{0}^{\beta}\!\!
	\left\{\hat{H}+\hat{a}^{\dagger}\eta(t)+\eta^{*}(t)\hat{a}\right\}\,dt
	\right]\vert z_{\mathrm{I}}\rangle.
\end{equation}
After dividing the time interval $\beta$ into $N$ segments, we express the T-product above as the limit
\begin{equation}
\begin{aligned}
	&\mathrm{T}\exp\left[-\int_{0}^{\beta}\!\!
	\left\{\hat{H}+\hat{a}^{\dagger}\eta(t)+\eta^{*}(t)\hat{a}\right\}\,dt
	\right]\\
	=&
	\lim\limits_{N\to\infty}
	\left\{\bm{1}-
	\epsilon(\hat{H}+\hat{a}^{\dagger}\eta_{N}+\eta^{*}_{N}\hat{a})\right\}
	\left\{\bm{1}-
	\epsilon(\hat{H}+\hat{a}^{\dagger}\eta_{N-1}+\eta^{*}_{N-1}\hat{a})\right\}
	\\
	&\hphantom{\lim\limits_{N\to\infty}}
	\times\cdots\times
	\left\{\bm{1}-
	\epsilon(\hat{H}+\hat{a}^{\dagger}\eta_{1}+\eta^{*}_{1}\hat{a})\right\}
\end{aligned}
\end{equation}
where we have written $\eta(t_{j})$ as $\eta_{j}$ for simplicity.
By making repeated use of the resolution of unity \eqref{eq:n-resolution01}, we obtain
\begin{equation}
\label{eq:nc-kernel}
\begin{aligned}
	&Z_{-}[\eta^{*},\,\eta]\\
	=&
	\lim\limits_{N\to\infty}
	\int\!\!\prod_{i=1}^{N-1}\frac{\ dz_{i}\,dz^{*}_{i}\ }{\pi}
	\exp\left[-\sum_{j=1}^{N-1}z^{*}_{j}z_{j}+
	\sum_{j=1}^{N}\left\{e^{-\epsilon}z^{*}_{j}z_{j-1}
	+\epsilon\left(z^{*}_{j}\eta_{j}-\eta^{*}_{j}z_{j-1}\right)\right\}
	\right]\\
	=&
	\exp\left[e^{-\beta}z^{*}_{\mathrm{F}}z_{\mathrm{I}}+
	\int_{0}^{\beta}\!\!
	\left\{z^{*}_{\mathrm{F}}e^{-(\beta-t)}\eta(t)-
	\eta^{*}(t)e^{-t}z_{\mathrm{I}}\right\}\,dt\right]\\
	&\times\exp\left[-
	\int\!\!\!\!\int_{0}^{\beta}\!\!
	\eta^{*}(t)\theta(t-t{'})e^{-(t-t{'})}\eta(t{'})\,dt\,dt{'}\right],
\end{aligned}
\end{equation}
where $\theta(t)$ denotes the step function.

We can compare this result with the corresponding one for a harmonic oscillator of the usual commutation relation $[\hat{a},\,\hat{a}^{\dagger}]=1$. The coherent state is defined by $\vert z\rangle=e^{\hat{a}^{\dagger}z}\vert0\rangle$ with its conjugate $\langle z\vert=\langle0\vert e^{z^{*}\hat{a}}$ for this case. Then the corresponding generating functional is given by
\begin{equation}
\label{eq:c-kernel}
\begin{aligned}
	Z_{+}[\eta^{*},\,\eta]=&\langle z_{\mathrm{F}}\vert
	\mathrm{T}\exp\left[-\int_{0}^{\beta}\!\!
	\left\{\hat{H}+\hat{a}^{\dagger}\eta(t)+\eta^{*}(t)\hat{a}\right\}\,dt
	\right]\vert z_{\mathrm{I}}\rangle\\
	=&
	\exp\left[e^{-\beta}z^{*}_{\mathrm{F}}z_{\mathrm{I}}-
	\int_{0}^{\beta}\!\!
	\left\{z^{*}_{\mathrm{F}}e^{-(\beta-t)}\eta(t)+
	\eta^{*}(t)e^{-t}z_{\mathrm{I}}\right\}\,dt\right]\\
	&\times\exp\left[+
	\int\!\!\!\!\int_{0}^{\beta}\!\!
	\eta^{*}(t)\theta(t-t{'})e^{-(t-t{'})}\eta(t{'})\,dt\,dt{'}\right].
\end{aligned}
\end{equation}
Note that the integral of the quadratic term of the external source has opposite sign in \eqref{eq:nc-kernel} compared to \eqref{eq:c-kernel} in addition to the difference in the sign of $z^{*}_{\mathrm{F}}$ coupled to $\eta(t)$.  
If we set $z^{*}_{\mathrm{F}}=z_{\mathrm{I}}=0$ in \eqref{eq:c-kernel}, the generating functional becomes the amplitude from the vacuum to the vacuum and we can define the generating functional of connected Green's functions by $W_{+}[\eta^{*},\,\eta]=-\log Z_{+}[\eta^{*},\,\eta]$. Explicitly, $W_{+}[\eta^{*},\,\eta]$ is given by
\begin{equation}
	W_{+}[\eta^{*},\,\eta]=-
	\int\!\!\!\!\int_{0}^{\beta}\!\!
	\eta^{*}(t)\theta(t-t{'})e^{-(t-t{'})}\eta(t{'})\,dt\,dt{'}.
\end{equation}
We now set $\eta^{*}(t)=\eta(t)=J(t)/\sqrt{2\,}$ to define
\begin{equation}
	W_{+}[J]=-\frac{\ 1\ }{\ 2\ }
	\int\!\!\!\!\int_{0}^{\beta}\!\!
	J(t)\Delta_{\mathrm{F}}(t-t{'})J(t{'})\,dt\,dt{'},
\end{equation}
where $\Delta_{\mathrm{F}}(t-t{'})$ is given by
\begin{equation}
	\Delta_{\mathrm{F}}(t)=\frac{\ 1\ }{\ 2\ }\left\{
	\theta(t)e^{-t}+\theta(-t)e^{t}\right\}
\end{equation}
and fulfills
\begin{equation}
	\left(-\frac{\ d^{2}\ \ }{\ dt^{2}\ }+1\right)
	\Delta_{\mathrm{F}}(t)=\delta(t).
\end{equation}
By defining $\varphi(t)$ by $\varphi(t)=\delta W_{+}[J]/\delta J(t)$ and making a Legendre transform, we find that the effective action $\varGamma_{+}[\varphi]$ is given by
\begin{equation}
\begin{aligned}
	\varGamma_{+}[\varphi]\equiv&
	W_{+}[J]-
	\int_{0}^{\beta}\!\!
	J(t)\varphi(t)\,dt\\
	=&\int_{0}^{\beta}\!\!
	\frac{\ 1\ }{\ 2\ }\left\{\left(\frac{\ d\varphi(t)\ }{\ dt\ }\right)^{2}+
	\varphi^{2}(t)\right\}\,dt,
\end{aligned}
\end{equation}
where use has been made of a relation
\begin{equation}
	J(t)=\left(\frac{\ d^{2}\ \ }{\ dt^{2}\ }-1\right)\varphi(t)
\end{equation}
for inversion.
Clearly, it is the action in the imaginary time for a harmonic oscillator. If we proceed the same way for the negative oscillator, we obtain the effective action 
\begin{equation}
\begin{aligned}
	\varGamma_{-}[\varphi]\equiv&
	W_{-}[J]-
	\int_{0}^{\beta}\!\!
	J(t)\varphi(t)\,dt\\
	=&-\int_{0}^{\beta}\!\!
	\frac{\ 1\ }{\ 2\ }\left\{\left(\frac{\ d\varphi(t)\ }{\ dt\ }\right)^{2}+
	\varphi^{2}(t)\right\}\,dt,
\end{aligned}
\end{equation}
where $W_{-}[J]$ is defined by $W_{-}[J]=-\log Z_{-}[J]$ after defining $Z_{-}[J]$ by setting $\eta^{*}(t)=\eta(t)=J(t)/\sqrt{2\,}$ in \eqref{eq:nc-kernel}.
Again, we find the action in the imaginary time for a negative oscillator here.
Although the action in the exponent of the time sliced path integral for the Feynman kernel has the opposite sign for the negative oscillator due to the imaginary eigenvalues of $\hat{p}$ and $\hat{q}$, the sign is recovered in the effective action. Let us state again: since the path integral is a definite integral and the action appears in its exponent is merely a part of the integrand, we cannot know everything at a glance from the integrand, i.e. the action of the path integral, without carrying out some evaluation of physical quantities such as the effective potential.

\end{document}